\documentclass{article}[12pt]
\usepackage{amsmath}
  \usepackage{paralist}
  \usepackage{graphics} 
  \usepackage{epsfig} 
 \usepackage[colorlinks=true]{hyperref}
\hypersetup{urlcolor=blue, citecolor=red}

\usepackage{graphicx}

\usepackage{psfrag}

\usepackage{epsfig}
\usepackage{amssymb}
\usepackage{amsthm}

 \numberwithin{equation}{section}

\newtheorem{theorem}{Theorem}[section]

\newtheorem{lemma}[theorem]{Lemma}
\newtheorem{proposition}[theorem]{Proposition}

\newtheorem{definition}[theorem]{Definition}
\newtheorem{remark}[theorem]{Remark}

\def\aepsilon{a}
\def\be{\begin{equation}}
\def \beq {\begin {eqnarray}}
\def \eeq {\end {eqnarray}}
\def \ba {\begin {eqnarray*}}
\def \ea {\end  {eqnarray*}}
\def \p {\partial}

\newcommand{\C}{{\Bbb C}}
\newcommand{\R}{{\Bbb R}}

\newcommand{\D}{{\Bbb D}}
\newcommand{\re}{{\rm Re \,}}
\newcommand{\im}{{\rm Im \,}}

\renewcommand{\Im}{{\rm Im \,}}

\newcommand{\av}[1]{\langle{#1}\rangle}

\newcommand{\OsmDb}{\Omega\setminus\Dbar}
\newcommand{\Cinf}{C^{\infty}}
\newcommand{\RR}{\mathbb{R}}
\newcommand{\CC}{\mathbb{C}}
\newcommand{\ke}{k_\text{e}}
\newcommand{\ki}{k_\text{i}}
\newcommand{\Dbar}{\overline{D}}
\newcommand{\Obar}{\overline{\Omega}}
\newcommand{\ubar}{\overline{u}}

\newcommand{\dO}{{\doo\Omega}}
\newcommand{\Hbar}{\overline{H}}
\newcommand{\dd}{\mathrm{d}}
\newcommand{\donu}{\doo_\nu}
\newcommand{\dor}{\doo_r}
\newcommand{\psido}{\psi\text{DO}}

\newcommand{\sm}{\setminus}
\newcommand{\doo}{\partial}

\newcommand{\Ga}{\Gamma}
\newcommand{\Gay}{\Gamma_1}
\newcommand{\Gak}{\Gamma_2}
\newcommand{\resgay}{\vert_{\Gamma_1}}
\newcommand{\resgak}{\vert_{\Gamma_2}}
\newcommand{\HeSo}{(\ref{Helmholtz})--(\ref{Sommerfeld}) }
\newcommand{\ee}{a_\text{e}}
\newcommand{\ei}{a_\text{i}}
\newcommand{\normi}[1]{\left\|#1\right\|}

\newcommand{\SkG}{S_k^{\Gamma}}
\newcommand{\SkGO}{S_k^{\Gamma,\Omega}}

\newcommand{\DkG}{D_k^{\Gamma}}
\newcommand{\DkGj}{D_k^{\Gamma_j}}
\newcommand{\DkGO}{D_k^{\Gamma,\Omega}}

\newcommand{\Nop}[2]{N_{{#1}}^{{#2}}}

\newcommand{\Kop}[2]{K_{{#1}}^{{#2}}}
\newcommand{\Vop}[2]{V_{{#1}}^{{#2}}}
\newcommand{\Dop}[2]{D_{{#1}}^{{#2}}}
\newcommand{\Sop}[2]{S_{{#1}}^{{#2}}}
\newcommand{\Top}[2]{T_{{#1}}^{{#2}}}

\newcommand{\KkG}{K_k^{\Gamma}}
\newcommand{\KkGj}{K_k^{\Gamma_j}}

\newcommand{\intxR}{\int_{|x|=R}}
\newcommand{\intdO}{\int_{\doo\Omega}}
\newcommand{\intdD}{\int_{\doo D}}
\newcommand{\intBsmO}{\int_{B_R(0)\setminus\Obar}}
\newcommand{\intD}{\int_D}
\newcommand{\intOsmD}{\int_{\Omega\setminus\Dbar}}
\newcommand{\mcAo}{\mathcal{A}_0}
\newcommand{\mcMo}{\mathcal{M}_0}
\newcommand{\mcAy}{\mathcal{A}_1}
\newcommand{\mcMy}{\mathcal{M}_1}

\newcommand{\mcAe}{\mathcal{A}_\eeta}
\newcommand{\mcMe}{\mathcal{M}_\eeta}

\newcommand{\AMo}{\mcAo + \mcMo}
\newcommand{\AMy}{\mcAy + \mcMy}

\newcommand{\AMe}{\mcAe + \mcMe}

\newcommand{\Sob}[2]{H^{{#1}}({#2})}
\newcommand{\pyve}[2]{\begin{pmatrix}{#1}\\{#2}\end{pmatrix}}

\newcommand{\udotu}[1]{u_{#1}\cdot\ubar_{#1}}

\newcommand{\AM}{\mathcal{A} + \mathcal{M}}

\newcommand{\CinfG}{C^\infty(\Gamma)}

\newcommand{\pheq}{\phantom{=}}
\newcommand{\OsmD}{\Omega\setminus\Dbar}
\newcommand{\dka}{\frac{d-2}{2}}

\newcommand{\Gk}{G_k}

\DeclareMathOperator{\ind}{\mathrm{ind}}
\DeclareMathOperator{\Real}{\mathrm{Re}}
\DeclareMathOperator{\Imag}{\mathrm{Im}}
\DeclareMathOperator{\supp}{\mathrm{supp}}

\newcommand{\eeta}{\eta}

\def \Box {$\square$}

\newcommand{\newtext}{\bf} 

\newcommand{\removed}[1]{}
\newcommand{\hiddenfootnote}[1]{}

\def\tilde{\widetilde}
\def \bfo {\begin {eqnarray*} }
\def \efo {\end {eqnarray*} }
\def \ba {\begin {eqnarray*} }
\def \ea {\end {eqnarray*} }
\def \beq {\begin {eqnarray}}
\def \eeq {\end {eqnarray}}
\def \supp {\hbox{supp}\,}

\def \re {\hbox{Re}\,}
\def \im {\hbox{Im}\,}

\def \ind {\hbox{Ind}\,}

\def \p {\partial}

\newcommand{\uvz}{\mathbf{u}_z}
\newcommand{\uvy}{\mathbf{u}_y}
\newcommand{\uvx}{\mathbf{u}_x}

\newcommand{\vr}{\mathbf{r}}

\newcommand{\vp}{\mathbf{p}}

\newcommand{\epr}{\varepsilon_\text{r}}

\newcommand{\mur}{\mu_r}

\newcommand{\epz}{\varepsilon_0}
\newcommand{\epsinf}{\varepsilon_\infty}

\newcommand{\vek}[1]{\mathbf{#1}}

\newcommand{\osder}[2]{\frac{\partial{#1}}{\partial{#2}}}

\title{On Absence and Existence of the Anomalous Localized Resonance without the Quasi-static Approximation}
\author{\scshape Henrik Kettunen, Matti Lassas, Petri Ola}
\date{}

\begin{document}
\maketitle

{\footnotesize
   \centerline{University of Helsinki}
 \centerline{Department of Mathematics and Statistics}
   \centerline{P.O.Box 68, 00014 University of Helsinki, Finland}
   \centerline{henrik.kettunen@helsinki.fi, matti.lassas@helsinki.fi, petri.ola@helsinki.fi}
}




\begin{abstract}
\noindent 
The paper considers the transmission problems for Helmholtz equation with bodies that have negative material
parameters. Such material parameters are used to model  metals on  optical frequencies
and so-called metamaterials.
As the absorption of the materials in the model tends to zero the fields may blow up.
When the speed of the blow up is suitable, this is
called the Anomalous Localized Reconance (ALR). In this paper
we study this phenomenon and formulate a new condition, the weak  Anomalous Localized Reconance (w-ALR),
where the speed of the blow up of fields may be slower.
Using this concept, we can study the blow up of fields in the presence of 
negative material parameters without the commonly used quasi-static approximation.
We give simple geometric conditions under which w-ALR or ALR may, or may not
appear. In particular, we show that in a case of a curved layer
of negative material with a strictly convex boundary neither ALR nor w-ALR 
appears with  non-zero frequencies (i.e. in the dynamic range) in dimensions $d\ge 3$.  In the case when the boundary of the negative
material contains a flat subset we show that the w-ALR always happens with some point sources
in dimensions $d\ge 2$. These results, together with the earlier  results of Milton et al. ( [22, 23])
and Ammari et al. ([2]) show that  for strictly convex bodies ALR
may  appear only for bodies so small that the quasi-static approximation is realistic.
This gives limits for size of the objects for which invisibility cloaking methods based on ALR
may be used.

\end{abstract}
\section{Introduction and statement of main results}

Consider a pair of bounded $\Cinf$-domains $D$ and $\Omega$ of $\RR^d$, $d \geq 2$, such that the closure of \(D\) is included in \(\Omega\). Given complex wave numbers $\ke$ and $\ki$, $\Im \ke, \,\Im \ki\geq 0$,
 we consider the  properties of the following transmission problem
\beq\label{Helmholtz} 
-(\Delta + \ke^2)v_1
& = & 0 \text{ in } D \\ \nonumber
-(\Delta+\ki^2)v_2 
& = & 0 \text{ in } \OsmDb \\
-(\Delta+\ke^2)v_3 & = & f \text{ in } \RR^d\sm\Omega, \quad f\in {\cal E}'(\RR^d\sm \Obar), \nonumber
\eeq
where on the {\em interior boundary} $\Ga_1 = \doo D$ we have the boundary conditions 
\begin{equation}\label{reunaa}
v_1\vert_{\Ga_1} = v_2\vert_{\Ga_1},\quad \tau_1\donu v_1 \vert_{\Ga_1} = \donu v_2 \vert_{\Ga_1}
\end{equation}
and on the {\em exterior boundary} $\Ga_2 = \dO$ we have
\begin{equation}\label{reunab}
v_2\vert_{\Ga_2} = v_3\vert_{\Ga_2}, \quad \tau_2\donu v_3\vert_{\Ga_2} = \donu v_2\vert_{\Ga_2}.
\end{equation}
Above, $\nu$ is the exterior unit normal vector of $\Omega\setminus \overline D$.
We also assume that the exterior field $v_3$ satisfies the (outgoing) Sommerfeld radiation condition at infinity, 
\beq\label{Sommerfeld}
\ke \neq 0, \, d\geq 2:& &v_3(x) = O\left(|x|^{2-d}\right), \quad (\doo_r - i\ke)v_3(x) = o\left(|x|^{2-d}\right),\\ \nonumber
& &\text{as } |x|\to\infty \text{ uniformly in } x/|x|  \in S^{d-1},\\ \nonumber
\ke = 0,\, d\geq 3: & &v_3(x) = O\left(|x|^{2-d}\right) \text{ as } |x|\to\infty \text{ uniformly in } x/|x| \in S^{d-1},\\ \nonumber
\ke = 0,\, d=2: & &v_3(x) = o\left(1\right) \text{ as } |x|\to\infty \text{ uniformly in } x/|x| \in S^{1},\\ \nonumber
\eeq
where $\p_r=\frac {x}{|x|}\,\cdotp \nabla$. Also, if \(d=2\) and \(k_e = 0\), we assume that the compactly supported source \(f\in {\cal E}'(\R ^2\setminus \overline \Omega)\) satisfies the vanishing condition 
\[
\langle f,1\rangle = 0.
\]

\begin{figure}[htbp]
\begin{center}\label{Fig-1}

\psfrag{1}{$\Omega$}
\psfrag{2}{$D$}
\includegraphics[width=5.5cm]{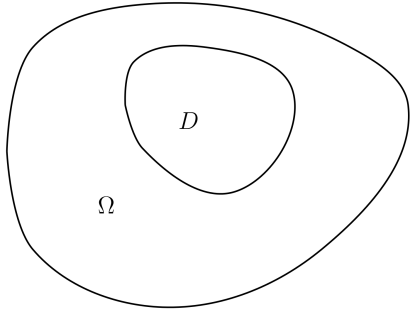}
\end{center}
\caption{Setting of the paper: Domain $\Omega\subset \R^d$ that contains the closure of domain $D$.
In the set $\Omega\setminus \overline D$ the material parameters approach negative value
and are positive outside this set.
}
 \end{figure}
We will also consider the equations (\ref{Helmholtz})-(\ref{reunab}) in divergence form.
To this end, we define
a piecewise constant function \(a_\eta\) by
\beq\label{Matti 1}
a_\eta (x)  = \ee > 0 \text{ in }D \text{ and } \RR^2\sm\Obar,\\
\label{Matti 2}
a_\eta (x)  = \ei = \ee(-1 + \eeta) \text{ in } \OsmDb, \quad \eeta \in \C,\ \im \eeta\geq 0,
\eeq 
and
\beq
\label{Matti 3}\tau_1 = \tau _2 = \tau = \frac{\ee}{\ei} = (-1+\eeta)^{-1}. 
\eeq
\noindent Typically, the parameter \(\eta\) above will be small and purely imaginary. A weak solution of 
\beq\label{div type eq.}
\nabla\cdot a_\eeta (x)\nabla u  + \omega^2(\chi_{D\cup \R ^d\setminus \overline \Omega} + b\chi_{\Omega\setminus \overline D})\mu _0  u= f \, {\rm in}\, \R ^d,\,  f\in {\cal E}'(\RR^d\sm \Obar),
\eeq
where \(b\) is a complex constant, is obtained from  \(u_1 = u|_D\), \(u_2 = u|_{\Omega\setminus\overline \D}\) and \(u_3 = u|_{\R^d\setminus\overline\Omega}\)  solving (\ref{Helmholtz})--(\ref{reunab}), where the transmission coefficients satisfy (\ref{Matti 3}). Note that since outside the interfaces  \(u\) solves a Helmholtz--equation, it has one sided weak normal derivatives on both interfaces.
Also, the wave numbers are determined by
\[
k_e^2 = \omega ^2 \mu_0 \ee ^{-1}, \, k_i^2=k_i(\eta)^2 := \omega ^2\mu_0 \ei ^{-1}b,
\]
and depending on our choice of \(b\) and \(\eeta\) the sign of \(\re k_i ^2\) may vary. 
We will in particular consider two physically interesting cases. In the first  case,
 \(b = 1\), and  \(\re k_i ^2 \leq 0\). In the second case,  \(b= -1\) and \(\re k_i ^2 \geq 0\). For more on the physical relevance of these cases, please see the Appendix at end of the article.
We also denote 
\beq
  k_{i,0}^2=k_i(\eta)^2|_{\eta=0} = -\omega ^2\mu_0 \ee ^{-1}b.
  \eeq

We are especially interested in the behavior of the solutions -- and of course in the unique solvability  -- as $\eeta \to 0$ when the ellipticity of \HeSo degenerates. Physically this corresponds to having a layer of (meta)material in $\OsmDb$. 
More precisely, as explained in the Appendix,  in $\R ^2$ this problem comes up when considering time-harmonic TE-polarized waves in the cylinder $\RR^2 \times \RR$ with the dielectric constants given by a piecewise constant \(a_\eta ^{-1}\). 

\medskip

\noindent It is known that in the case when $\Omega=B(0,R_1)$ and $D=B(0,1)$
are discs
(see \cite{ACKLM,ACKLMII,Milton05,Milton06}) and $\omega=0$ that when $\eeta \to 0$ there is a limit radius $R^*>0$ s.t. if 
$$\supp(f) \subset (\RR^2\sm\Obar)\cap\{x;|x|>R^*\}$$
the solution of \HeSo will have a bounded $H^1$-norm in $\OsmDb$ as $\eeta \to 0$, but when 
$$\supp(f)\subset (\RR^2\sm\Obar)\cap B(0,R^*)$$
the $H^1(\Omega\setminus \overline D)$--norm of $u_2$ blows up at least  as \({\cal O}(|\eeta|^{-1/2})\). 
This phenomenon is called {\em anomalous localized resonance} (ALR).  To clarify the results of this paper, we make the following formal definitions:

\begin{definition}
Let \(v_i^{\eta}\), \(i=1,2,3\),  be the unique solutions of \HeSo for \(\eta \not = 0\) with a given, fixed source term \(f\in {\cal E}'(\R ^d \setminus \overline \Omega)\).
\begin{enumerate}
\item If \(\limsup_{\eta \to 0} \| v_2^\eta \|_{H^1(\Omega\setminus\overline D)} = \infty\), we say that the {\em Weak Anomalous Localized Resonanace} (w-ALR) occurs.

\item If \(\| v_2^\eta \|_{H^1(\Omega\setminus\overline D)}  \geq C/|\eta|^{1/2}\) as \(\eta \to 0\), we say that the {\em Anomalous Localized Resonanace} (ALR) occurs.
\end{enumerate}

\end{definition}

\noindent In this paper we show that neither ALR nor w-ALR happens in $\RR^d$, $d\geq 3$, when the boundaries of $\Omega$ and $D$ are strictly convex as embedded hypersurfaces of \(\R ^d\). We also prove that if the exterior boundary has a flat part then w-ALR will occur even without the quasi-static approximation. 
Numerical simulations explained in the Appendix support the hypothesis that w-ALR is a weaker phenomenon than ALR. Note, that in \cite{ACKLM} the authors define a condition called weak-CALR. In our case this is equivalent to having
\[
\limsup_{\eta \to 0}|\eta|^{1/2} \| v_2^\eta \|_{H^1(\Omega\setminus\overline D)} = \infty
\] 
and hence is stronger than w-ALR.

\medskip

In the seminal  papers by Milton et al  \cite{Milton05,Milton06} it was observed that ALR happens
in the two-dimensional case when $\Omega$ and $D$  are co-centric disc, i.e., $\Omega\setminus  \overline D$ is an annulus,
in the quasi-static regime.
This case corresponds to a ``perfect lens'' made of negative material with a small conductivity $|\eta|$  when
$|\eta|\to 0$. When this device  is located in a homogeneous electric field and a
polarizable point-like object is taken close to the object, the point-like object produces 
a point source due to the background field. When the object is sufficiently close to the annulus, the 
induced fields in the annulus blow up as $|\eta|\to 0$.
Surprisingly, the fields in the annulus create a field which far away cancels the field produced by the point like-object.
This result can be interpreted by saying that the annulus makes the point-like object invisible. Presently, this  
phenomena is called ``exterior cloaking''. It is closely related to other type of invisibility cloaking techniques,  
the transformation optics based cloaking, see \cite{GKLU1,GKLUrev1,GKLUrev2,GKLU-pnas,GLU2,GLU3,KSVW,Kohn2,PSS1} and active cloaking, see
\cite{Vasquez1,Vasquez2}. These cloaking examples can be considered as counterexamples 
for unique solvability of various inverse problems that show the limitations of various imaging modalities.
\cite{U2,U3}.

Results of Milton et al  \cite{Milton05,Milton06} raised plenty of interest and motivated many  studies on the topic.
The cloaking due to anomalous localized resonance is studied in the quasi-static regime for a general domain in
 \cite{ACKLM}. There, it is shown that in $\R^2$ the resonance happens for a large class of the sources
 and that the resonance occurs not because of system approaching an eigenstate, but 
  because of the divergence of an infinite sum of terms related to spectral decomposition of the Neumann-Poincare operator.
  In \cite{Ketal} the ALR is studied in the quasi-static regime in the two dimensional case
 when the outer domain $\Omega$ is a disc and the core $D$ is an arbitrary domain compactly supported in $\Omega.$
 ALR in the case of confocal ellipses is studied in
 \cite{Chung}.

 In  \cite{ACKLMII}, it was shown that the cloaking due to anomalous localized resonance does not happen
 in $\R^3$ when  $\Omega$ and $D$  are co-centric balls.  In \cite{ACKLM3},  cloaking due to anomalous localized resonance
 is connected to transformation optics and there it is shown that ALR may happen in three dimensional case when the coefficients of the
 equations are appropriately chosen matrix-valued functions, i.e. correspond to the non-homogeneous anisotropic material.

Earlier, ALR has been studied without the quasi-static approximation both in the 2 and 3 dimensional cases 
 in  \cite{Nguyen1,Nguyen2}. In these papers the appearance of ALR is connected to the compatibility of the sources. 
 The compatibility  means that for these sources there exists solutions for certain non-elliptic boundary value
 problems, that are analogous to the so-called interior transmission problems.

 In this paper we show that w-ALR either happens or does not happen when certain simple geometric conditions hold:
 We show that w-ALR - and hence also ALR - do not happen in 3 and higher dimensional case when the boundaries $\partial \Omega$ and $\partial D$ are strictly
 convex, and also show that  w-ALR does happen in $d$-dimensional case, $d\geq 2$ with some sources 
when the boundary of $\partial \Omega$ contains a flat part. These results show that w-ALR is directly related to geometric
properties of the boundaries, and that the behavior of the solutions of the Helmholtz equation with a positive
frequency is very different to the solutions in the quasi-static regime.




The first main result deals with the solvablity of the case \(\tau _1 = \tau _2 = \tau = -1\) when the ellipticity of the transmission problem degenerates. Below we will use the notation \(\Hbar^{1}_\text{loc}(\RR^d\sm\Obar) =\{u;\, u= w|_{\R^3\setminus \overline \Omega}, \, w\in H^1_{\rm loc} (\R ^d)\}\). Also, we assume in both Theorems below that the following injectivity assumption is valid:

\begin{itemize}
\item
{\bf Injectivity assumption (A)}: Assume that the equation \HeSo has only the trivial solution $v_1= v_2 = v_3 = 0$ when  \(f=0\).
\end{itemize}

The first main result shows that under certain geometric conditions on the boundary interfaces the limit problem \(\eta = 0\) is solvable.

\begin{theorem} \label{tokathm}
Let $d\geq 3$. Assume that the interior boundary \(\Gamma _1\) and the exterior boundary \(\Gamma _2\) are smooth and strictly convex. 
Assume also that 
$\eeta=0$, so that  $\tau = -1$, and that
$\ke > 0$, $k_{i,0}^2$ is not a Dirichlet eigenvalue of $-\Delta$  in \(D\) and \(\R^d \setminus \overline \Omega\), and \(k_e^2\)  is not a Dirichlet eigenvalue of \(-\Delta\) in \(\Omega\). 
Then given \(K\subset \RR ^d \setminus \overline \Omega\) compact and \(f\in H^{s} (\RR ^d \setminus \overline \Omega)\) with \(\supp (f) \subset K\), the problem \HeSo has a unique solution $v_1 \in H^{1}(D)$, $v_2 \in H^{1}(\Omega\sm D)$ and $v_3 \in \Hbar^{1}_\text{loc}(\RR^d\sm\Obar)$ such 
$$
\normi{v_2}_{H^{1}(\OsmDb)} \leq C_K \normi{f}_{H^{s}(\RR^d\sm\Obar)}.
$$
\end{theorem}

We can also prove the following limiting result when \(\eeta \to 0\).

\begin{theorem}\label{stability} Let $d\geq 3$.
Assume that the interfaces  \(\Gamma _1\) and  \(\Gamma _2\) are smooth and strictly convex and let
\(f\) be as in the previous Theorem.
Let $\tau = \tau (\eeta)  = ( -1  + \eeta)^{-1}\), $\eeta\in \C$.
Assume also that $\ke > 0$, $k_{i,0}^2$ is not a Dirichlet eigenvalue of $-\Delta$  in \(D\), and \(k_e^2\)  is not a Dirichlet eigenvalue of \(-\Delta\) in \(\Omega\).
Then the problem \HeSo with $\tau  = \tau(\eeta)$ is uniquely solvable for \(|\eta|\) small enough, and  if $v_i^\eeta$, $i=1,2,3$ are its solutions and \(v_i\), \(i =1,2,3\)  the solutions given by Theorem \ref{tokathm}, we have as $\eeta \to 0$ in a complement of a open conical neighborhood of \(i\R\),$$v_1^\eeta \xrightarrow[H^{1}(D)]{} v_1, \quad v_2^\eeta \xrightarrow[H^{1}(\OsmDb)]{} v_2,
\quad v_3^\eeta \xrightarrow[\Hbar_\text{loc}^{1}(\RR^d\sm\Obar)]{} v_3.$$ 
\end{theorem}
\medskip

\noindent Both of these theorems will be proven in section four of this paper.
\medskip

\noindent
Some comments are in order. First of all, under the assumptions of the above theorems, given a fixed source distribution \(f\) supported in \(\R ^d\setminus \overline\Omega\), the solution \(v_2^\eta\) tends to a \(H^1\)--function \(v_2\) as \(\eta \to 0\), and thus there is no blow up.
\medskip

\noindent Secondly, in Proposition 4.1 we give sufficient conditions for the injectivity assumption (A) to hold. Especially, if the wave numbers come from a divergence type equation with piecewise constant coefficients, so that (\ref{Matti 1}) -- (\ref{Matti 3}) are valid, the injctivity will hold.
\medskip

\noindent Finally, the remaining conditions assumed of the wavenumbers are related to the boundary integral equation we use, and quarantee the equivalence of it with the original transmission problem. We believe that using another reduction to the boundary these could be relaxed. 



\bigskip

\section{Layer potentials}
As a first step we are going to reduce \HeSo to an equivalent problem on the boundary interfaces. The first step it to replace the source \(f\) with equivalent boundary currents. So, fix \(f\in H^{s}_0 (\RR ^d \setminus \overline \Omega)\) and let \(v\) be the unique solution of the problem
\beq\label{reduktio}
-(\Delta + k_e^2) v & = & f \, \, \mbox{in \(\RR ^d \setminus \overline \Omega\)}, \\
v|_{\Gamma _2} & = & 0
\eeq
that satisfies the Sommerfeld radiation condition (\ref{Sommerfeld}).
Then, if we let \(\tilde v _3 = v_3  - v\) in \HeSo we see that \(v_1,\, v_2\) and \(\tilde v_3\) will satisfy the transmission problem
consisting of equation (\ref{Helmholtz}) with radiation condition
(\ref{Sommerfeld}) and the transmission conditions
%
\begin{equation}\label{reunaa BBB}
v_1\vert_{\Ga_1} = v_2\vert_{\Ga_1},\quad \tau_1\donu v_1 \vert_{\Ga_1} = \donu v_2 \vert_{\Ga_1}
\end{equation}
and
\[
\tilde v_3\resgak = v_2\resgak - f_2, \quad \tau_2\donu \tilde v_3\resgak =  \donu v_2\resgak -  g_2
\]
where \(f_2 =  v\resgak \in H^{1/2}(\Gamma _2)\) and \(g_2 = \donu v\resgak \in H^{-1/2}(\Gamma _2)\). Hence, we are going to study first the transmission problem 
\beq\label{transmissio-ongelma}
-(\Delta + \ke^2)u_1 & = & 0 \text{ in } D 
\\
-(\Delta+\ki^2)u_2  & = & 0 \text { in } \OsmDb 
\\
-(\Delta+\ke^2)\tilde u_3 & = & 0 \text{ in } \RR^d\sm\Obar 
\eeq
\beq
u_1\resgay = u_2\resgay - f_1, \quad \tau_1\donu u_1\resgay & = & \donu u_2\resgay - g_1  
\\
u_2\resgak = u_3\resgak-f_2, \quad \tau_2\donu u_3\resgak &= & \donu u_2\resgak - g_2 
\eeq
\begin{equation}\label{radiaatio}
u_3 \text{ satisfies Sommerfeld condition (\ref{Sommerfeld}).} 
\end{equation}
\smallskip

\noindent where  \(f_1 \in H^{s}(\Gamma _1)\), \(g_1 \in H^{s-1}(\Gamma _1)\), \(f_2  \in H^{s}(\Gamma _2)\) and \(g_2  \in H^{s-1}(\Gamma _2)\) for some real value of \(s\). Notice also that since the 
source \(f\) is supported away from \(\Gamma _2\) the solution \(v\) of (\ref{reduktio}) will actually be smooth near \(\Gamma _2\), and hence 
the boundary jumps $f_i$ and $g_i$ will be \(C^\infty\) functions. This will be crucial for our argument.
\medskip

The reduction to the boundary will be done in the usual way, i.e. by using suitable layer potential operators.  Given $k \in \CC$, $\Imag k \geq 0$, let 
$$
\Gk(x) =
\begin{cases}
\frac{i}{4}\left(\frac{k}{2\pi|x|}\right)^{\dka}H^{(1)}_{\dka}(k|x|), \quad k\neq 0, \quad d \geq 2\\[2ex]
\frac{\Ga\left(\dka\right)}{4\pi^{\frac{d}{2}}}|x|^{2-d}, \quad k = 0, \quad d \geq 3\\[2ex]
\frac{1}{2\pi}\ln |x|, \quad d=2, \, k=0.
\end{cases}
$$
be the fundamental solution of $-(\Delta+k^2)$ in $\RR^d$ 
satisfying Sommerfeld condition (\ref{Sommerfeld}).

\medskip

\noindent For $\Ga = \Ga_1$ or $\Ga = \Ga_2$, we define using these kernels the (volume) single layer operators by 
$$\SkG(\phi)(x) = \int_{\Ga}\Gk(x-y)\phi(y)\dd S(y), \quad x \notin \Ga, \quad \phi \in C^\infty(\Ga)$$
Sometimes when we wish to emphasize that we are restricting these operators to a domain $\Omega$, $\Ga \cap \Omega = \emptyset$, we use the notations $\SkGO(\phi)
=\SkG(\phi)|_{\Omega}$. These operators define continuous mappings $\SkG:\CinfG \to D'(\RR^d\sm\Ga)$. 
\medskip

\noindent Similarly, we define the (volume) double layer operators by 
$$\DkG(\phi)(x) = \int_{\Ga}\frac{\doo \Gk(x-y)}{\doo \nu(y)}\phi(y)\dd S(y), \quad x\notin \Ga,
\quad \phi \in \CinfG$$
where $\nu$ will always denote the exterior unit normal to $\Omega\sm D$. Like for the single layer potentials, we will occasionally denote the restrictions of these to \(\Omega\subset \R^d \setminus \Gamma\) by $\DkGO$.  Also, $\DkG: \Cinf(\Gamma) \to D'(\RR^d\sm \Ga)$ continuously.
\medskip

\noindent Mapping properties of these operators between appropriate Sobolev spaces are also well known (see \cite{BdM}, \cite{McL} or \cite{RS}, page 156, Theorem 4): For all $s\in \RR$ we have $\SkGO:H^s(\Ga)\to H^{s+\frac{3}{2}}(\Omega)$ and $\DkGO:H^s(\Ga)\to H^{s+\frac{1}{2}}(\Omega)$ if $\Omega \subset \RR^d\sm \Ga$ is a  bounded domain.
\medskip

We need traces of these operators on both \(\Gamma _1\) and \(\Gamma _2\). Hence, let \(\gamma ^{+}_{0,j}\) be the trace operator on \(\Gamma_j\) from the complement of \(\Omega\setminus \overline D\), 
that is, $\gamma ^{+}_{0,j}(u)= u|_{\Gamma_j}$
for $u\in H^1((\R^d\setminus \overline\Omega)\cup  D)$.
Respectively let \(\gamma ^{-}_{0,j}\) be the trace--opearator on \(\Gamma_j\) from \(\Omega\setminus \overline D\),
that is, $\gamma ^{-}_{0,j}(u)=u|_{\Gamma_j}$
for $u\in H^1(\Omega\setminus \overline D)$.
Then, for \(\phi \in H^{s}(\Gamma _j)\), \(s>-1\), we have
\begin{equation}\label{trace}
\gamma ^{+}_{0,j} S^{\Gamma _j} _{k} \phi = V_k^{\Gamma _j} \phi = \gamma ^{-}_{0,j} S^{\Gamma _j} _{k} \phi,
\end{equation}
where 
\begin{equation}
V_k^{\Gamma _j} \phi(x)  = \int_{\Gamma _j} \Gk(x-y)\phi(y)\dd S(y)
\end{equation}
is the trace-single-layer operator on $\Gamma _j$. 
\medskip

\noindent
Also, if  \(\psi \in H^{s}(\Gamma _j)\), \(s>0\), for the traces of the double layer we have the jump relations
\begin{equation}\label{double-}
\gamma ^{-}_{0,j} \DkGj\psi + \frac{\psi}{2} = \gamma ^{+}_{0,j}\DkGj\psi - \frac{\psi}{2} = \KkGj\psi,
\end{equation}
where $\KkGj$ is the trace-double-layer operator on \(\Gamma _j\) given by
\begin{equation}
\KkG\phi(x) = p.v.\int_{\Gamma _j}\frac{\doo\Gk(x-y)}{\doo \nu(y)}\phi(y)\dd S(y).
\end{equation}

\noindent
The maps \(K_{k}^{\Gamma_j}: H^s(\Gamma _j) \to H^s(\Gamma _j)\) and \(S_{k}^{\Gamma_j}: H^s(\Gamma _j) \to H^{s+1}(\Gamma _j)\) are continous pseudodifferential operators for any \(s\in \R\).
\medskip

Next, let \(\gamma ^{+}_{1,j}\) be the trace of the normal derivative on \(\Gamma_j\) from the complement of \(\Omega\setminus \overline D\), 
that is, $\gamma ^{+}_{1,j}(u)=\p_\nu u|_{\Gamma_j}$
for $u\in H^1((\R^d\setminus \overline\Omega)\cup  D)$.
Respectively let \(\gamma ^{-}_{1,j}\) be the trace of the normal derivative on \(\Gamma_j\) from \(\Omega\setminus \overline D\),
that is, $\gamma ^{-}_{1,j}(u)=\p_\nu u|_{\Gamma_j}$
for $u\in H^1(\Omega\setminus \overline D)$.
. For the normal derivatives of the single layer potentials we have the jump relations
\begin{equation}\label{normal-trace}
\gamma ^{-}_{1,j} S_k^{\Gamma_j}\phi - \frac{\phi}{2} = \gamma ^{+}_{1,j} S_k^{\Gamma_j}\phi + \frac{\phi}{2} = K_k^{*,\Gamma_j}\phi,
\end{equation}
for any \(\phi \in H^{-1/2}(\Gamma _j)\), where the operator \(K_k^{*,\Gamma_j}\) is the adjoint trace-double-layer operator on \(\Gamma _j\) given by
\begin{equation}
K_k^{*,\Gamma_j}\phi (x)  = p.v.\int_{\Gamma _j}\frac{\doo\Gk(x-y)}{\doo \nu(x)}\phi(y)\dd S(y),
\end{equation}
\medskip

For \(H^1\)--solutions of an inhomogeneous Helmholz--equation with an 
\(L^2\)--source one can define normal traces weakly using Green's theorems. With this interpretation, for any \(\psi \in H^{1/2}(\Gamma _j)\),  we also have the traces
\begin{equation}
\gamma ^{-}_{1,j} D_k^{\Gamma_j}\psi = \gamma ^{+}_{1,j} D_k^{\Gamma_j}\psi = N_k^{\Gamma_j}  \psi,
\end{equation}
where the hypersingular integral operator $N_k^{\Gamma_j}$ has (formally) the kernel 
\[
\frac{\doo^2 \Gk(x-y)}{\doo\nu(x)\doo\nu(y)}, \, \, x,\, y\in \Gamma_j, \, x\not = y.
\]
The maps \(K_{k}^{*,\Gamma_j}: H^s(\Gamma _j) \to H^s(\Gamma _j)\) and \(N_{k}^{\Gamma_j}: H^s(\Gamma _j) \to H^{s-1}(\Gamma _j)\) are continous pseudodifferential operators for any \(s\in \R\).
\bigskip

\section{Reduction to the boundary}
We will follow  the ideas of \cite{K-M} adapted to our situation, where we have two interfaces instead of just one.
Let us consider (\ref{transmissio-ongelma})--(\ref{radiaatio}). Write an ansaz for $u_1$ and $u_3$:
\begin{equation}\label{ansatzu1}
u_1 = \Sop{\ke}{\Gay,D}(\phi), \quad \phi \in H^{-\frac{1}{2}}(\Ga_1),
\end{equation}
\begin{equation}\label{ansatzu3}
u_3 = \Sop{\ke}{\Gak,\RR^d\sm\Obar}(\psi),\quad \psi \in H^{-\frac{1}{2}}(\Ga_2), 
\end{equation}
and to
$$u_2 \in {\cal L}:=\{v\in H^1(\OsmDb);\ -(\Delta+\ki^2)v=0\}$$
we apply the representation theorem (see for example \cite{CK})
to get
\beq
u_2 &=& \Sop{\ki{}}{\Gay,\OsmDb}\left(\left.\frac{\doo u_2}{\doo\nu}\right\vert_{\Ga_1}\right) 
- \Dop{\ki{}}{\Gay,\OsmDb}\left(u_2\vert_{\Ga_1}\right)\\
& &+ \Sop{\ki{}}{\Gak,\OsmDb}\left(\left.\frac{\doo u_2}{\doo\nu}\right\vert_{\Ga_2}\right)
- \Dop{\ki{}}{\Gay,\OsmDb}\left(u_2\vert_{\Ga_2}\right),
\text{ in }\OsmDb.
\nonumber \eeq
Taking traces from $D$ on $\Gay$ we get
\begin{equation}\label{traceD}
\gamma ^{+}_{0,1}u_1 = \Vop{\ke}{\Gay}(\phi), \quad  \gamma ^{+}_{1,1}u_1 = \Kop{\ke}{*,\Gay}(\phi) - \frac{\phi}{2},
\end{equation} 
and taking traces from $\RR^d\sm\Obar$ on $\Ga_2$ we get
\begin{equation}\label{traceDD}
\gamma ^{+}_{0,2}u_3 = \Vop{\ke}{\Gak}(\psi), \quad \gamma ^{+}_{1,2}u_3 = \Kop{\ke}{*,\Gak}(\psi) - \frac{\psi}{2}.
\end{equation}
Now, denote
\ba
A_if &= \left.\Sop{\ki{}}{\Ga_j,\OsmDb}(f)\right\vert_{\Ga_i}, \quad i\neq j, \quad f \in H^{-\frac{1}{2}}(\Ga_j)\\[1ex]
B_if &= \left.\frac{\doo \Sop{\ki{}}{\Ga_j,\OsmDb}(f)}{\doo\nu}\right\vert_{\Ga_i},
\quad i\neq j, \quad f \in H^{-\frac{1}{2}}(\Ga_j)\\[1ex]
R_ig &= \left.\Dop{\ki{}}{\Ga_j,\OsmDb}(g)\right\vert_{\Ga_i}, \quad i\neq j, \quad g \in H^{\frac{1}{2}}(\Ga_j)\\[1ex]
S_ig &= \left.\frac{\doo\Dop{\ki{}}{\Ga_j,\OsmDb}(g)}{\doo\nu}\right\vert_{\Ga_i},\quad i\neq j, \quad g \in H^{\frac{1}{2}}(\Ga_j). 
\ea
Note that all these operators have \(C^\infty\)--smooth kernels. 
Taking traces of $u_2$ on $\Ga_1$ and $\Ga_2$ we get
\beq
\hspace{-5mm}\gamma ^{-}_{0,1}u_2 &= \Vop{\ki{}}{\Gay}\left(\donu u_2\resgay\right) - \Kop{\ki{}}{\Gay}\left(u_2\resgay\right)
+\left.\frac{u_2}{2}\right\vert_{\Ga_1} + A_1\left(\donu u_2\resgak\right) - R_1(u_2\resgak)\label{traceeka}\\
\hspace{-5mm}\gamma ^{-}_{0,2}u_2 &= \Vop{\ki{}}{\Gak}\left(\donu u_2\resgak\right) - \Kop{\ki{}}{\Gak}\left(u_2\resgak\right)
+\left.\frac{u_2}{2}\right\vert_{\Ga_2} + A_2\left(\donu u_2\resgay\right) - R_2(u_2\resgay)\label{tracetoka}
\eeq
and for the traces of the normal derivatives we get
\begin{equation}
\gamma ^{-}_{1,1} u_2 = \Kop{\ki{}}{*,\Gay}(\donu u_2\resgay) + \left.\frac{\donu u_2}{2}\right\vert_{\Ga_1}
-\Nop{\ki{}}{\Gay}(u_2\resgay)
+ B_1\left(\donu u_2\resgak\right) - S_1(u_2\resgak)
\nonumber
\end{equation}
and
\begin{equation}
\gamma ^{-}_{1,2}u_2 = \Kop{\ki{}}{*,\Gak}(\donu u_2\resgak) + \left.\frac{\donu u_2}{2}\right\vert_{\Ga_2}
-\Nop{\ki{}}{\Gak}(u_2\resgak)
+ B_2\left(\donu u_2\resgay\right) - S_2(u_2\resgay).
\nonumber
\end{equation}
\vspace{1ex}
Recall next  the transmission conditions
\beq
u_1\resgay &= u_2\resgay - f_1, \quad \tau_1\donu u_1\resgay = \donu u_2\resgay - g_1,\label{transeka}\\
u_3\resgak &= u_2\resgak - f_2, \quad \tau_2\donu u_3\resgak = \donu u_2\resgak - g_2\label{transtoka},
\eeq
$$f_i \in H^{\frac{1}{2}}(\Ga_i), \quad g_i \in H^{-\frac{1}{2}}(\Ga_i)$$

\noindent If one substitutes (\ref{transeka})  to  (\ref{traceeka}) one gets an integral equation
$$
\frac{\gamma ^{+}_{0,1}u_1}{2} + \Kop{\ki{}}{\Gay}(\gamma ^{+}_{0,1}u_1) - \tau_1\Vop{\ki{}}{\Gay}(\gamma ^{+}_{1,1} u_1)
+ \tau_2A_1(\gamma ^{+}_{1,2}u_3) - R_1(\gamma ^{+}_{0,2}u_3) = \tilde{f}_1
$$
where
\begin{equation}\label{reuna-arvot1}
\tilde{f}_1:= -\frac{f_1}{2} + \Vop{\ki{}}{\Gay}(g_1) - \Kop{\ki{}}{\Gay}(f_1) + A_1(g_2) - R_1(f_2).
\end{equation}
Next, using (\ref{traceD}) and  (\ref{traceDD}) we write the boundary values of \(u_1\) and \(u_3\) in terms of \(\psi\) and \(\phi\). This gives  an boundary integral equation on \(\Gamma _1\):
\begin{equation}\label{yht323}
\frac{1}{2}\left(\Vop{\ke}{\Gay} + \tau_1\Vop{\ki{}}{\Gay}\right)(\phi)
+\left(\Kop{\ki{}}{\Gay}\Vop{\ke}{\Gay} - \tau_1\Vop{\ki{}}{\Gay}\Kop{\ke}{*,\Gay}\right)(\phi)
+ M_1(\psi) = \tilde{f}_1,
\end{equation}
where
$$
M_1(\psi) = \tau_2A_1\Kop{\ke}{*,\Gak}(\psi) - \frac{\tau_2}{2}A_1(\psi) - R_1\Vop{\ke}{\Gak}(\psi)
$$
is a smoothing operator. Substituting similarly (\ref{transtoka})  to  (\ref{tracetoka}) one gets an boundary integral equation 
on \(\Gamma _2\):
\begin{equation}\label{yht324}
\frac{1}{2}\left(\Vop{\ke}{\Gak}+\tau_2\Vop{\ki{}}{\Gak}\right)(\psi)
+\left(\Kop{\ki{}}{\Gak}\Vop{\ke}{\Gay}-\tau_2\Vop{\ki{}}{\Gak}\Kop{\ke}{*,\Gak}\right)(\psi)
+M_2(\phi) = \tilde{f}_2,
\end{equation}
where
$$
M_2(\phi) = \frac{\tau_2}{2}A_2(\phi) - \tau_2 A_2\Kop{\ke}{*,\Gay}(\phi) + R_2\Vop{\ke}{\Gay}(\phi)
$$
is smoothing and
\begin{equation}\label{reuna-arvot2}
\tilde{f}_2:= -\frac{f_2}{2} + \Vop{\ki{}}{\Gak}(g_2) - \Kop{\ki{}}{\Gak}(f_2) + A_2(g_1) - R_2(f_1).
\end{equation}
\medskip

\noindent We combine the two equations above as follows:  for $(\phi,\psi) \in H^{-\frac{1}{2}}(\Ga_1)\times H^{-\frac{1}{2}}(\Ga_2)$ we have
\begin{equation}\label{Matriisiyht}
(\AM)
\begin{pmatrix}\phi \\  \psi \end{pmatrix}
= \tilde{f}:= \begin{pmatrix}\tilde{f}_1 \\ \tilde{f}_2 \end{pmatrix}
\end{equation}
where
\beq\label{Matriisiyht2}
\mathcal{A}&= &\begin{pmatrix}
\mathcal{A}_1 & 0 \\ 0 & \mathcal{A}_2
\end{pmatrix},\\
\mathcal{M} &=& \begin{pmatrix}
0 & M_1 \\ M_2 & 0
\end{pmatrix}
\nonumber
\eeq
and
\ba
\mathcal{A}_1 &= \frac{1}{2}\left(\Vop{\ke}{\Gay} + \tau_1\Vop{\ki{}}{\Gay}\right)
+\left(\Kop{\ki{}}{\Gay}\Vop{\ke}{\Gay} - \tau_1\Vop{\ki{}}{\Gay}\Kop{\ke}{*,\Gay}\right),\\
\mathcal{A}_2 &= \frac{1}{2}\left(\Vop{\ke}{\Gak}+\tau_2\Vop{\ki{}}{\Gak}\right)
+\left(\Kop{\ki{}}{\Gak}\Vop{\ke}{\Gay}-\tau_2\Vop{\ki{}}{\Gak}\Kop{\ke}{*,\Gak}\right).
\ea
This is the integral equation we are going to study. 
\medskip

The next proposition establishes conditions under which (\ref{Matriisiyht}) is equivalent with the original transmission problem (\ref{transmissio-ongelma})--(\ref{radiaatio}).

\begin{proposition}\label{prop33}
Assume $\ki^2$ is not an Dirichlet eigenvalue of \(-\Delta\) in \(D\) or in \(\R^3 \setminus {\overline \Omega}\) . If $u_1 \in H^1(D)$, $u_2 \in H^1(\OsmDb)$ and $u_3 \in \overline{H}^1_\text{loc}(\RR^d\sm\Obar)$ solve (\ref{transmissio-ongelma})--(\ref{radiaatio}), then $\phi\in \Sob{-\frac{1}{2}}{\Ga_1}$, $\psi \in \Sob{-\frac{1}{2}}{\Ga_2}$ satisfying (\ref{ansatzu1})--(\ref{ansatzu3}) solve (\ref{Matriisiyht}).

Conversely, assume that  $(\phi,\psi) \in \Sob{-\frac{1}{2}}{\Ga_1}\times\Sob{-\frac{1}{2}}{\Ga_2}$ solve (\ref{Matriisiyht}). Define $u_1$ and $u_3$ by (\ref{ansatzu1})--(\ref{ansatzu3}) and $u_2$ by
\beq\label{u2yhtalo}
u_2 &=& \Sop{\ki{}}{\Gay}\left(\tau_1\left[\Kop{\ke}{*,\Gay}(\phi) - \frac{\phi}{2}\right]+g_1\right)
-\Dop{\ki{}}{\Gay}\left(\Vop{\ke}{\Gay}(\phi) + f_1\right)\\
& &\pheq+\Sop{\ki{}}{\Gak}\left(\tau_2\left[\Kop{\ke}{*,\Gak}(\psi) - \frac{\psi}{2}\right]+g_2\right)
-\Dop{\ki{}}{\Gak}\left(\Vop{\ke}{\Gak}(\psi) + f_2\right).
\nonumber
\eeq
Then the triplet $(u_1,u_2,u_3)\in \Sob{1}{D}\times\Sob{1}{\OsmDb}\times\overline{H}^1_\text{loc}(\RR^d\sm\Obar)$ will solve (\ref{transmissio-ongelma})--(\ref{radiaatio}).
\end{proposition}

\proof It only remains to prove the second claim. So define \(u_2\) by (\ref{u2yhtalo}) and $u_1$ and $u_3$ by (\ref{ansatzu1})--(\ref{ansatzu3}) where $\phi\in \Sob{-\frac{1}{2}}{\Ga_1}$, $\psi \in \Sob{-\frac{1}{2}}{\Ga_2}$  solve (\ref{Matriisiyht}). By taking traces on \(\Gamma _1\) and \(\Gamma _2\) we immediately recover the transimission conditions 
\[
u_1\resgay = u_2\resgay - f_1, u_3\resgak = u_2\resgak - f_2.
\]
To prove the transmission conditions for the normal derivatives we define for  \(x \in D \cup \RR^d\sm\Obar\),
\beq
v(x) &=& \Sop{\ki{}}{\Gay}\left(\tau_1\left[\Kop{\ke}{*,\Gay}(\phi)-\frac{\phi}{2}\right]+g_1\right)(x)
-\Dop{\ki{}}{\Gay}\left(\Vop{\ke}{\Gay}(\phi)+f_1\right)(x)\\
 \nonumber &&\pheq+
\Sop{\ki{}}{\Gak}\left(\tau_2\left[\Kop{\ke}{*,\Gak}(\psi)-\frac{\psi}{2}\right]+g_2\right)(x)
-\Dop{\ki{}}{\Gak}\left(\Vop{\ke}{\Gak}(\psi)+f_2\right)(x).
\nonumber
\eeq
Then using (\ref{Matriisiyht}) we see that \(v\) solves  
\begin{equation*}
\begin{cases}
(\Delta + \ki^2)v = 0 \text{ in } D\\
v\resgay = 0
\end{cases}
\end{equation*}
and since we assumed that \(k_i^2\) was not an Dirichlet eigenvalue of \(-\Delta\) in \(D\), we get that \(v = 0\) in \(D\). Similarily the restriction of \(v\) to \(\R^3 \setminus {\overline \Omega}\) is
a solution of 
\begin{equation*}
\begin{cases}
-(\Delta + \ki^2)v = 0 \text{ in } \RR^d\sm\Obar\\
v\vert_{\dO} = 0
\end{cases}
\end{equation*}
satisfying Sommerfeld condition (\ref{Sommerfeld}).
Hence by the assumptions, \(v = 0\) also in \(\R^3 \setminus {\overline \Omega}\). Taking traces of the normal derivative of \(v\) from \(D\) we 
get
\ba
0 &=& \gamma ^{+}_{1,1} v 
=\Kop{\ki{}}{*,\Gay}\left(\tau_1\left[\Kop{\ke}{*,\Gay}(\phi)-\frac{\phi}{2}\right]+g_1\right)\\
&&\pheq-\frac{1}{2}\left(\tau_1\left[\Kop{\ke}{*,\Gay}(\phi)-\frac{\phi}{2}\right]+g_1\right)
-\Nop{\ki{}}{\Gay}\left(\Vop{\ke}{\Gay}(\phi)+f_1\right)\\
& &\pheq+B_1\left(\tau_2\left[\Kop{\ke}{*,\Gak}(\psi)-\frac{\psi}{2}\right]+g_2\right)
-S_1\left(\Vop{\ke}{\Gak}(\psi)+f_2\right),
\nonumber
\ea
or equivalently,
\beq\label{Yht333}
\tau_1\left[\Kop{\ke}{*,\Gay}(\phi)-\frac{\phi}{2}\right] + g_1
&=&\Kop{\ki{}}{*,\Gay}\left(\tau_1\left[\Kop{\ke}{*,\Gay}(\phi)-\frac{\phi}{2}\right]+g_1\right)\\
\nonumber
& &\hspace{-3cm}\pheq+\frac{1}{2}\left(\tau_1\left[\Kop{\ke}{*,\Gay}(\phi)-\frac{\phi}{2}\right]+g_1\right)
-\Nop{\ki{}}{\Gay}\left(\Vop{\ki{}}{\Gay}(\phi)+f_1\right)\\
 & &\hspace{-3cm}\pheq+B_1\left(\tau_2\left[\Kop{\ke}{*,\Gak}(\psi)-\frac{\psi}{2}\right]+g_2\right)
-S_1\left(\Vop{\ke}{\Gak}(\psi)+f_2\right).
\nonumber
\eeq
The left-hand side of (\ref{Yht333}) is $\tau_1\donu u_1 + g_1$. The right-hand side is equal to $\donu u_2\resgay$. Hence we have shown the second equation in (\ref{transeka}). Proceeding similarly, but taking traces of \(v\) from \(\R^3 \setminus {\overline \Omega}\), we get the second equation of (\ref{transtoka}). \($\Box$\)

\begin{remark}
Note that the Dirichlet-spectrum of $-\Delta$ on $D$ is discrete and positive, so that always, if $\ki^2 \leq 0$ or if  $\Imag \ki^2 \not = 0$, we have the first condition. Similarly, $\Imag \ki \geq 0$, $\ki \not = 0$, is enough to guarantee that the second assumption of Proposition ~\ref{prop33} is valid. In the case of most interest to us we have \( \ke^2 = \omega^2\ee\mu_0 >0\) and \(k_i^2 = \omega^2\ei\mu_0\) with \(\ei = \ee(-1  + \eeta)\), where \(\ee\) and and imaginary part of \(\eeta\) are positive. Hence the assumptions of Proposition ~\ref{prop33} are valid.
\end{remark}
\bigskip

\section{Absence of ALR}
As the first step in proving the solvability and stability when \(\eeta \to 0\) we give a standard uniqueness result (see \cite{CS}):

\begin{proposition}\label{prop41}
Assume $\ke>0$, $\Real\left( \ki^2/\tau_2\right)\geq0$ and $\Real \tau_1\tau_2^{-1}\ke^2 \geq 0$. Then the problem \HeSo has at most one solution. Also, if the wave numbers are associated to a divergence type equation, i.e. (\ref{Matti 1}) -- (\ref{Matti 3}) are satisfied with a real valued parameter \(b\), the uniqueness holds.
\end{proposition}

\proof Let $R>0$ be so large that $\Obar\subset B_R(0)$. Then the Sommerfeld Radiation Condition implies
\begin{equation}\label{farfield}
\intxR\dor\udotu{3}\dd S =  ik_e \lim _{R\to \infty} \intxR|u_3|^2 \, \dd S + o(1) \quad \text{as}Ê\,R\to \infty.
\end{equation}
Now using Green's formula
\ba
&&\intxR\dor u_3\cdot\ubar_3\dd S - \intdO\donu \udotu{3} \dd S\\
&&=\intBsmO\Delta \udotu{3} \dd x = \intBsmO -\ke^2|u_3|^2\dd x,
\ea
i.e.
$$
\intxR\dor\udotu{3}\dd S = -\ke^2\intBsmO|u_3|^2\dd x + \intdO\donu \udotu{3}\dd S.
$$
Now, if $u_1$, $u_2$ and $u_3$ solve the homogeneous version of \HeSo,
$$
\intdO\donu\udotu{3}\dd S = \intdO \tau_2^{-1}\donu\udotu{2}\dd S
$$
and
\begin{equation*}
\intdO\donu\udotu{2}\dd S + \intdD\donu\udotu{2}\dd S\\ 
= \intOsmD\Delta\udotu{2}\dd x = -\intOsmD \ki^2|u_2|^2\dd x,
\end{equation*}
so that
\ba
\intxR\dor\udotu{3}\dd S & = & -\ke^2\intBsmO|u_3|^2\dd x - \frac{1}{\tau_2}\intOsmD \ki^2|u_2|^2\dd x\\
& - & \frac{1}{\tau_2}\intdD\donu\udotu{2}\dd S.
\ea
Finally,
\begin{equation*}
\intdD\donu\udotu{2}\dd S = \tau_1\intdD\donu\udotu{1}\dd S\\
=-\tau_1\intD -\ke^2|u_2|^2\dd x = \tau_1 \ke^2\intD |u_1|^2\dd x, 
\end{equation*}
and thus
\beq\label{green-relaatio}
&&\intxR\donu\udotu{3}\dd S\\ \nonumber
&&=-\ke^2\intBsmO |u_3|^2\dd x - \frac{1}{\tau_2}\intOsmD \ki^2|u_2|^2\dd x
-\frac{\tau_1}{\tau_2}\ke^2\intD |u_1|^2\dd x.
\eeq
Taking real parts of (\ref{green-relaatio}) we get using (\ref{farfield}) in the first case that
\begin{equation}
\lim_{R\to\infty} -\ke^2\intBsmO |u_3|^2\dd x
-\intOsmD \Real\left\{\tau_2^{-1}\ki^2\right\}|u_2|^2\dd x
-\Real\left(\frac{\tau_1}{\tau_2}\ke^2\right)\intD |u_1|^2\dd x =0.
\end{equation}
Every term on the left hand side is nonpositive, so we must have
$$
0 = \lim_{R\to \infty}\intBsmO |u_3|^2\dd x = \intOsmD |u_2|^2 \Real\left\{\tau_2^{-1}\ki^2\right\}\dd x
=\Real\left(\frac{\tau_1}{\tau_2}\ke^2\right) \intD |u_1|^2\dd x,
$$
hence $u_1 = u_2 = u_3 = 0$. In the second case we have 
\[
\tau_1 = \tau _2 = \tau = \frac{\ee}{\ei} = (-1+\eeta)^{-1}
\]
and \(k_e^2 = \omega ^2 \mu_0 \ee ^{-1}, \, k_i^2= \omega ^2\mu_0 \ei ^{-1}b\), so that
\[
\frac{1}{\tau_2}k_i^2 = \omega ^2\mu_0 \ei ^{-1}b(-1 + \eeta) =  \omega ^2 \mu_0 \ee ^{-1}b.
\]
Hence all the integrals in the right hand side of (\ref{green-relaatio}) are real, and by taking imaginary parts we get
\[
ik_e \lim _{R\to \infty} \intxR|u_3|^2 \, \dd S = 0,
\]
so Rellich's theorem implies \(u_3= 0\). Thus by the transmission conditions  the Cauchy-data of \(u_2\) vanishes on the exterior boundary, so Holmgren's uniqueness theorem implies \(u_2 = u_1 = 0\).  $\qed$


It is well known  (see for example \cite{CS} or \cite{TII}) that on a smooth compact surface \(\Gamma\) without boundary the single-layer potentials $\Vop{k}{\Ga}$ are classical pseudodifferential operators \((\psi DO\)'s) of order $-1$ with principal symbol 
$$
\sigma_{(-1)}\left(\Vop{k}{\Ga}\right)(x, \xi') = c_d|\xi'|, \quad
\xi'\in \Top{*}{*}(\Ga), \quad x\in\Ga,
$$
and that \(4\Nop{k}{\Ga}\)  is the parametrix of $\Vop{k}{\Ga}$, so that
$$
\sigma_{(1)}\left(\Nop{k}{\Ga}\right)(x,\xi') = c_d^{-1}|\xi'|^{-1}/4, \quad \xi'\in\Top{*}{*}(\Ga)\sm\{0\}, \quad x\in\Ga.
$$
Also -- and this is important to us -- even though formally of order 0, the double-layer and its adjoints are in fact of order $-1$, and hence compact as operators $\Sob{s}{\Ga} \to \Sob{s}{\Ga}$. For the principal symbol of \(\Kop{k}{*,\Ga}\) we have (see \cite{O} or \cite{TII}, Proposition C.1, p.453) 
$$
\sigma_{(-1)}(\Kop{k}{*,\Ga})(x',\xi')
= {a_d}d_\Ga(x')|\xi'|^{-3}\left(l_x(\xi',\xi')-\sum_j\lambda_j(x')|\xi'|^2\right),
$$
where \(a_d\) is a nonzero constant,  $d_{\Gamma}(x')$ is the density of the surface measure on $\Ga$, $l_x$ is the second fundamental form of $\Ga$ (embedded in $\RR^d$) and $\lambda_j(x')$  are the principal curvatures of $\Ga$, i.e. eigenvalues of $l_x$. Hence, if $\Ga$ is strictly convex, $\Kop{k}{*,\Ga}$ is an elliptic operator of order $-1$.

\begin{proposition}\label{fredholm}
Assume $\tau_1,\tau_2 \neq -1$. Then the integral operator $\AM$ defined by (\ref{Matriisiyht})--(\ref{Matriisiyht2}) is elliptic $\psido$ of order $-1$, and hence a Fredholm operator
\begin{equation*}
\AM: 
\begin{matrix} \Sob{s}{\Gay} \\ \oplus \\ \Sob{s}{\Gak} \end{matrix}
\to \begin{matrix} \Sob{s}{\Gay} \\ \oplus \\ \Sob{s}{\Gak} \end{matrix}
\end{equation*}
for all $s \in \RR$. Also, ${\rm ind}\,\mathcal{A}=0$.
\end{proposition}

\proof The principal symbol of $\AM$ is
\begin{equation*}
\begin{pmatrix}
\frac{1}{2}(1+\tau_1)|\xi'|^{-1} & 0 \\
0 & \frac{1}{2}(1+\tau_1)|\xi'|^{-1}
\end{pmatrix}
\end{equation*}
proving the ellipticity. Also, if $k=0$, then $\Vop{0}{\Ga}$ is self-adjoint on $\Ga$, and hence $\ind\left(\Vop{0}{\Ga}\right)=0$. Since $\Vop{k}{\Ga}-\Vop{0}{\Ga}$ is of order $<-1$, and $\Kop{k}{\Ga_j}$ and $\Kop{k}{*,\Ga_j}$ are of order $-1$,
\begin{equation*}
\ind(\AM) = \ind\begin{pmatrix}
\Vop{0}{\Gay} & 0 \\ 0 & \Vop{0}{\Gak} \end{pmatrix} = 0. \qed
\end{equation*}
\medskip

\noindent From now on we only consider the case $d\geq 3$.

\begin{lemma} For the difference of single layer potentials we have
$$
\Vop{\ke}{\Ga_j} - \Vop{\ki{}}{\Ga_j} \in \Psi_{cl}^{-3+\epsilon}(\Ga_j)\quad\hbox{for all}\quad \epsilon > 0.
$$
\end{lemma}

\proof  
Now, if $d \geq 3$,
\ba
G_k(x) & = & \frac{i}{4}\left(\frac{k}{2\pi|x|}\right)^{\dka}H^{(1)}_{\dka}(k|x|)\\
& = &
\begin{cases}
C_d|x|^{-d+2} + O\left([k|x|]^{-d+4}\right), \quad d = 3,5,6,\ldots\\[2ex]
C_4|x|^{-2} + k\tilde{C}_4\ln\left(\frac{k|x|}{2}\right) + O(k|x|), \quad d=4,
\end{cases}
\ea
where the constant  $C_d$ is independent of $k$, and in fact we have full asymptotic expansions for the remainder in terms of power $(k|x|)^{-d+2\nu}$, $\nu = 2,3,\ldots$. Hence,
\begin{equation*}
\begin{cases}
\Vop{k_1}{\Ga_j} - \Vop{k_2}{\Ga_j}
\in\Psi_{cl}^{-3}(\Ga_j), \quad d=3,5,6,\ldots\\[2ex]
\Vop{k_1}{\Ga_j} - \Vop{k_2}{\Ga_j} \in 
\Psi_{cl}^{-3+\epsilon}(\Ga_j), \quad d=4
\end{cases}
\end{equation*}
for all $\epsilon > 0$. \, \($\Box$\)
\medskip

We can now show:
\begin{proposition}\label{degenfredholm}
Assume $\tau_1=\tau_2=-1$ and that $\Gay$ and $\Gak$ are strictly convex smooth hypersurfaces of $\RR^d$ with \(d\geq 3\). Then $(\AM)$ is an elliptic $\psido$ of order $-2$ with index {\rm 0}.
\end{proposition}

\proof  As $\tau_1=\tau_2=-1$,
\begin{equation*}
\mathcal{A}=
\begin{pmatrix}
\Kop{0}{\Gay}\Vop{0}{\Gay} + {\Vop{0}{\Gay}\Kop{0}{*,\Gay}} & 0\\
0 & \Kop{0}{\Gak}\Vop{0}{\Gak} + \Vop{0}{\Gak}\Kop{0}{*,\Gak}
\end{pmatrix}\mod \Psi_{\rm cl}^{-3+\epsilon}
\end{equation*}

\noindent Since $\Kop{0}{\Ga_j}\Vop{0}{\Ga_j} + \Vop{0}{\Ga_j}\Kop{0}{*,\Ga_j}$ is self-adjoint, $\ind(\AM)=0$. Also, by Calderon's identities (see \cite{CS})
$$
\Kop{0}{\Ga_j}\Vop{0}{\Ga_j} = \Vop{0}{\Ga_j}\Kop{0}{*,\Ga_j}
$$
and hence the principal symbol of $\mathcal{A}$ is
\begin{equation*}
Cd_{\Gay}(x')|\xi'|^{-4}\left(
\begin{array}{ll} 
a_1 (x',\xi') & 0\\
0 & a_2(x',\xi')
\end{array}
\right)
\end{equation*}
where 
\[
a_k(x',\xi') = l_{x'}^{\Gamma _k}(x',\xi')-\sum_{j=1}^{d-1} \lambda_j^{\Gamma _k}(x')|\xi'|^2,
\]
 $l_x^{\Gamma _k}$ is the second scalar fundamental form of $\Gamma _k$, and $\lambda_j^{\Gamma _k}$ are the principal curvatures  of $\Gamma _k$, i.e., eigenvalues of $l^{\Ga_k}$. If $\Ga_k$ is strictly convex, then for all \(x\in\Ga_k\), $\lambda^{\Gamma_k}_j (x)$ are either positive or negative, and since $\lambda_j^k$ are eigenvalues of $l_x^{\Gamma _k}(\xi',\xi')$, $l_{x'}^{\Gamma _k}(\xi',\xi')-\sum\lambda_j^{\nu}(x')|\xi'|^2$ is correspondingly either negative or positive definite, so $\AM$ is elliptic of order $-2$. \qed
\medskip

Next we consider the unique solvability of the boundary integral equation. We start by proving the uniquenes, and for this we make no additional assumptions on \(\tau_1\) and \(\tau_2\).

\begin{lemma}\label{boundaryuniq}
Assume that the conditions on the wavenumbers \(k_e\) and \(k_i\) of  Propositions \ref{prop41} and \ref{prop33} hold, \(k_e^2 \) is not an Dirichlet eigenvalue of \(\Omega\),  and
$$
(\AM)\pyve{\phi}{\psi} = 0, \quad \pyve{\phi}{\psi} \in \Sob{s}{\Gay}\times\Sob{s}{\Gay}.
$$
Then $\phi_1=\phi_2=0$.
\end{lemma}

\proof Assume 
$$
(\AM)\pyve{\phi}{\psi} = 0.
$$
Then by Proposition \ref{prop33}
\begin{equation*}
\begin{cases}
u_1 = \Sop{\ke}{\Gay,D}(\phi)\\[2ex]
u_2 = \Sop{\ki{}}{\Gay}\left(\tau_1\left[\Kop{\ke}{*,\Gay}(\phi)-\frac{\phi}{2}\right]\right)
-\Dop{\ki{}}{\Gay}\Vop{\ke}{\Gay}(\phi)\\[1ex]
\phantom{u_2 = }+\Sop{\ki{}}{\Gak}\left(\tau_2\left[\Kop{\ke}{*,\Gak}(\psi)-\frac{\psi}{2}\right]\right)
-\Dop{\ki{}}{\Gak}\Vop{\ke}{\Gak}(\psi), \quad \text{ in } \OsmD\\[2ex]
u_3 = \Sop{\ke}{\Gak,\RR^3\sm\Obar}(\psi)
\end{cases}
\end{equation*}
will solve \HeSo with $f=0$, so by Proposition \ref{prop41} we have  $u_1=0$, $u_2=0$ and $u_3=0$. Hence
$$
0 = u_1\resgay = \Vop{\ke}{\Gay}(\phi), \quad 0 = u_3\resgak = \Vop{\ke}{\Gak}(\psi) = 0.
$$
Also
\begin{eqnarray}
0 &= \donu u_1\resgay = \Kop{\ke}{*,\Gay}(\phi) - \frac{\phi}{2}, \label{star equation}\\
0 &= \donu u_3\resgak = \Kop{\ke}{*,\Gak}(\psi) - \frac{\psi}{2}. \nonumber
\end{eqnarray}
Define now
\begin{eqnarray*}
\tilde{u}_1 & = &\Sop{\ke}{\RR^d\sm\Dbar}(\phi) \text{ in }\RR^3\sm D,\\
\tilde{u}_2 & = &\Sop{\ke}{\Omega}(\psi) \text{ in }\Omega.
\end{eqnarray*}
Then
\begin{eqnarray*}
\tilde{u}_1\resgay & = & \Vop{\ke}{\Gay}(\phi) = 0\\
\tilde{u}_2\resgak &  = & \Vop{\ke}{\Gak}(\psi) = 0.
\end{eqnarray*}
Since the exterior Dirichlet problem is always uniquely solvable if $\ke > 0$  (see \cite{CK}), $\tilde{u}_1 \equiv 0$ in $\RR^d\sm \overline D$, so by taking traces of $\donu\tilde{u}_1$ on $\Gay$ we get
\begin{equation}
0 = \Kop{\ke}{*,\Gay}(\phi) + \frac{\phi}{2}. \label{star equation prime}
\end{equation} 
Thus $(\ref{star equation})$ and $(\ref{star equation prime})$ imply that $\phi=0$. Similarly, since $\ke ^2$ is not an interior Dirichlet eigenvalue of $\Omega$, we get $\psi = 0$. $\qed$
\medskip

\noindent  Combining Lemma \ref{boundaryuniq} and Propositions \ref{fredholm} and \ref{degenfredholm} we now get
\begin{proposition}\label{prop410}
Assume again that the conditions on the wavenumbers \(k_e\) and \(k_i\) of  Propositions \ref{prop41} and \ref{prop33} hold, and that \(k_e ^2\) is not an Dirichlet eigenvalue of \(\Omega\). 
Let
$$
\pyve{\tilde f_1}{\tilde f_2} \in \begin{matrix} \Sob{s}{\Gay} \\ \oplus \\ \Sob{s}{\Gak} \end{matrix},
$$
where \(s>-1\). 
Then if either
\medskip

a) $\tau_1,\tau_2 \neq -1$ ,
\medskip

\noindent 
or 
\medskip

b) $\tau_1 = \tau_2 = -1$, \(d\geq 3\)  and $\doo D$ and $\dO$ are strictly convex, 
\medskip

\noindent
the boundary integral equation (\ref{Matriisiyht}) has a unique solution
$$
\pyve{\phi_1}{\phi_2} \in \begin{matrix} \Sob{s-1}{\Gay} \\ \oplus \\ \Sob{s-1}{\Gak} \end{matrix}\, \text{in case a)}
\text{ or }
\pyve{\phi_1}{\phi_2} \in \begin{matrix} \Sob{s-2}{\Gay} \\ \oplus \\ \Sob{s-2}{\Gak} \end{matrix},\, \text{in case b)}
\text{ respectively. } \qed
$$
\end{proposition}

\begin{remark} 
Notice that if \(\tilde f_1\) and \(\tilde f_2\) are given by (\ref{reuna-arvot1}) and (\ref{reuna-arvot2}) respectively, where \(f_i\) and \(g_i\) are determined by the source \(f\in H^{s}(\RR ^d \setminus \overline \Omega)\) with a compact support contained in \(\R^d \setminus \overline \Omega\) as described at the beginning of the section 2, then \(\tilde f_1\) and \(\tilde f_2\) will be smooth functions and hence the above proposition holds with any \(s >-1\). Hence especially the field \(u_2\)  will belong to \( H^1 (\Omega\setminus \overline D)\), and we have proven Theorem \ref{tokathm}.


\end{remark}

\medskip

To prove Theorem \ref{stability} we need the following result:

\begin{proposition}\label{prop411}
Assume that the conditions on the wavenumbers \(k_e\) and \(k_i\) of  Propositions \ref{prop41} and \ref{prop33} hold, and that \(k_e\) is not an Dirichlet eigenvalue of \(\Omega\).  Assume also that $\doo D$ and $\dO$ are strictly convex. Let 
$$
\tau_1 = \tau_2 = \tau(\eeta):= (-1+\eeta)^{-1}, \quad \eeta \in \CC\setminus \Ga, \quad \eeta \neq 0,
$$
where $\Ga$ a conic neighbourhood of $i\RR$. Given
$$
\pyve{\tilde f_1}{\tilde f_2} \in \begin{matrix} \Sob{s}{\Gay} \\ \oplus \\ \Sob{s}{\Gak} \end{matrix}, \, s>-1,
$$
let
$$
\pyve{\phi_1(\eeta)}{\phi_2(\eeta)} \in \begin{matrix} \Sob{s-1}{\Gay} \\ \oplus \\ \Sob{s-1}{\Gak} \end{matrix}
$$
be the unique solution of (\ref{yht323}) with $\tau_1=\tau_2 = \tau(\eeta)$. Also, let
$$
\pyve{\phi_1}{\phi_2} \in \begin{matrix} \Sob{s-2}{\Gay} \\ \oplus \\ \Sob{s-2}{\Gak} \end{matrix}
$$
be the unique solution of (\ref{Matriisiyht}) with $\tau_i = \tau_e = -1$. Then as $\eeta \to 0$ in \(\C\setminus\Sigma\), we have 
$$
\pyve{\phi_1(\eeta)}{\phi_2(\eeta)} \to \pyve{\phi_1}{\phi_2} 
$$
in the  the space \(\Sob{s-2-\rho}{\Gay} \oplus \Sob{s-2-\rho}{\Gak}\) with all positive \(\rho\).
\end{proposition}
\medskip

Before the proof we give the following lemma:

\begin{lemma}\label{lemma412}
Let $M$ a compact Riemannian manifiold  and  \(E\) a smooth hermitian vector bundle on \(M\). Assume that  $P_0 = P_0(x,D): C^\infty  (M,E)\to C^\infty(M,E)$ is an elliptic $\psido$ on $M$ of order $m$ and $P_1 = P_1(x,D): C^\infty (M,E)\to C^\infty(M,E)$ is an invertible elliptic $\psido$ of order $(m-1)$. Assume also that $\varepsilon P_0(x,D) + P_1(x,D)$ is invertible for
$\varepsilon \in \CC$ with  $0 < |\varepsilon|$ being small enough. Finally, assume that there exists a cone \(\Sigma_0 \subset \CC\) and \(\eeta _0>0\) such that
$$
\Real \left\{\varepsilon P_{0,m}(x,\xi)P_{1,m-1}(x,\xi)^{-1}\right\} \geq C|\varepsilon| |\xi|, \quad |\xi| \text{ large enough},
\quad |\varepsilon| < \eeta, \quad \varepsilon \in \Sigma_0.
$$
Then for \(|\xi| \text{ large enough}\), \(\varepsilon \in \Sigma_0\) and \(|\varepsilon| < \eeta\),   we have uniform bound
\begin{equation*}
\normi{P_1(\varepsilon P_0 + P_1)^{-1}u}_{\Sob{s}{M}} \leq C_s\normi{u}_{\Sob{s}{M}}, \, u \in \Sob{s}{M}.
\end{equation*}
\noindent Here $P_{\nu,l}(x,\xi)$ is the principal symbol of $P_\nu(x,D)$ of order $l$ ($l=m$ for $P_0$, $l=m-1$ for $P_1$).
\end{lemma}

\proof The proof is based on  G\aa rding's inequality, see for example \cite{TII}.  Let
\begin{equation}
Q_{\varepsilon} (x,D) = \left(\varepsilon P_0 (x,D) + P_1 (x,D)\right)P_1^{-1}(x,D)
= \varepsilon S(x,D) +id,
\end{equation}
where \(S(x,D) = P_0(x,D)P_1^{-1}(x,D).\)
Given $f\in \Cinf(M,E)$ we compute
\ba
\normi{Q_\varepsilon f}^2_{L^2(M,E)} &=& (Q_\varepsilon^*Q_\varepsilon f, f)_{L^2(M,E)}\\
&=&\left((\overline{\varepsilon}S^*(x,D)+id)(\varepsilon S(x,D)+id)f,f\right)_{L^2(M,E)}\\
&=&|\varepsilon|^2\normi{S(x,D)f}^2_{L^2(M,E)} 
+ 2\left(\Real\{\varepsilon S(x,D)\}f,f\right)_{L^2(M,E)} + \normi{f}^2_{L^2(M,E)}.
\ea
Now when $|\varepsilon| < \eeta$, $\varepsilon \in \Sigma_0$, in local coordinates we have
$$
\Real \varepsilon S(x,\xi) \geq C|\varepsilon| |\xi|, \quad  \text{ for \(|\xi|\) large enough}.
$$
Hence G\aa rding's inequality gives, for any \(s\in \R\),
\begin{equation*}
\normi{Q_\varepsilon f}^2_{L^2(M,E)} \geq C_0|\varepsilon| \normi{f}^2_{\Sob{\frac{1}{2}}{M,E}}
+ \normi{f}^2_{L^2(M,E)} - C_s|\varepsilon| \normi{f}^2_{\Sob{s}{M,E}}
\geq \frac{1}{2}\normi{f}_{L^2(M,E)}
\end{equation*}
by taking  \(\varepsilon\) is small enough and \(s=0\). Hence, if
$$
f = P_1(x,D)\left(\varepsilon P_0(x,D)+P_1(x,D)\right)^{-1}u, \quad u\in\Cinf(M,E),
$$
we get
$$
\normi{P_1(x,D)\left(\varepsilon P_0(x,D)+P_1(x,D)\right)^{-1}u}_{L^2(M,E)} 
\leq 2\normi{u}_{L^2(M.E)}, \quad |\varepsilon|<\eeta, \quad \varepsilon\in\Sigma_0.
$$
Let \(\Lambda_s=(I-\Delta_g)^{2/2}: \Sob{s}{M,E} \to L^2(M,E)\) be an isomorphism with principal symbol \( \Lambda_S(x,\xi) = \av{\xi}^S\).  Here,
$\Delta_g$  is the Laplace-Beltrami operator on $M$.
Then
applying this estimate to
\begin{eqnarray*}
\tilde{P}_1(x,D) &= \Lambda_S(x,D)P_1(x,D),\\ 
\tilde{P}_0(x,D) &= \Lambda_S(x,D)P_0(x,D), 
\end{eqnarray*}
we get (note that $S(x,D) = \tilde{P}_0(x,D)\tilde{P}_1(x,D)^{-1}$),
\ba
&&\normi{\tilde{P}_1(x,D)\left(\varepsilon \tilde{P}_0(x,D)+\tilde{P}_1(x,D)\right)^{-1}\tilde{u}}_{L^2(M,E)}\\
&=&\normi{\Lambda_S(x,D)P_1(x,D)
\left(\varepsilon P_0(x,D)+P_1(x,D)\right)^{-1}\Lambda_{-S}\tilde{u}}_{L^2(M,E)}\\
&\leq& 2\normi{\tilde{u}}_{L^2(M,E)} \quad \hbox{for all} \quad \tilde{u}\in\Cinf(M,E)
\ea
or, in terms of \(u = \Lambda_S\tilde{u}\),
\begin{equation*}
\normi{P_1(x,D)\left(\varepsilon P_0(x,D)+P_1(x,D)\right)^{-1}u}_{\Sob{S}{M,E}}
\leq C_S\normi{u}_{\Sob{S}{M,E}},
\end{equation*}
with \(C_S\) independent of  \(\varepsilon, \quad |\varepsilon| < \eeta, \quad \varepsilon\in\Sigma_0. \qed\)
\medskip

\begin{remark} The principal symbols of $V_k$ and $K_k$ are (see \cite{CK}, \cite{O} and \cite{TII}),
\begin{eqnarray*}
\sigma_{V_k}(x,\xi) &  = & C_d|\xi'|^{-1}\\
\sigma_{\Kop{k}{*}}(x,\xi) & =  & a_dd_\Ga(x')|\xi'|^{-3}
\left(l_{x'}(\xi',\xi')-\sum_j\lambda_j(x')|\xi'|^2\right)
\end{eqnarray*}
with $C_d$ and $a_d$ dimension dependent constants and $d_\Ga(x')$ is the density of the surface measure on $\Ga$. Since $\sigma_{V_k} = \sigma_{V_0}$ and $V_0$ is self-adjoint, the coefficient $C_d$ must be real. Also, by the identity
$$
K_kV_k = V_kK_k^*
$$
we have
$$
\sigma_K(x,\xi) = \sigma_{K^*}(x,\xi) = \overline{\sigma_{K}(x,\xi)},
$$
implying that $a_d$ is real.
\end{remark}
\medskip

\noindent {\bf Proof of  Proposition ~\ref{prop411}}:
With the obvious notation
\begin{eqnarray*}
\pyve{\phi_1(\eeta)}{\phi_2(\eeta)} &= (\AMe)^{-1}\tilde{f},\\[1ex]
\pyve{\phi_1}{\phi_2} &= (\AMo)^{-1}\tilde{f}
\end{eqnarray*}
where $\mcAe$ is an elliptic diagonal $\psido$ of the form
\begin{equation*}
\mcAe =
\begin{pmatrix}
\mathcal{A}_{\eeta,1} & 0\\
0 & \mathcal{A}_{\eeta,2}
\end{pmatrix}
\end{equation*}
with
\begin{eqnarray*}
\mathcal{A}_{\eeta,1} &= \frac{1}{2}\left(\Vop{\ke}{\Gay}+(-1+\eeta)\Vop{\ki{}}{\Gay}\right)
+\left(\Kop{\ki{}}{\Gay}\Vop{\ke}{\Gay}-(-1+\eeta)\Vop{\ki{}}{\Gay}\Kop{\ke}{*,\Gay}\right),\\
\mathcal{A}_{\eeta,2} &= \frac{1}{2}\left(\Vop{\ke}{\Gak}+(-1+\eeta)\Vop{\ki{}}{\Gak}\right)
+\left(\Kop{\ki{}}{\Gak}\Vop{\ke}{\Gak}-(-1+\eeta)\Vop{\ki{}}{\Gak}\Kop{\ke}{*,\Gak}\right),
\end{eqnarray*}
and
\begin{equation*}
\mcAo =
\begin{pmatrix}
\mathcal{A}_{0,1} & 0\\
0 & \mathcal{A}_{0,2}
\end{pmatrix}
\end{equation*}
where
\begin{eqnarray*}
\mathcal{A}_{0,1} &= \frac{1}{2}\left(\Vop{\ke}{\Gay}-\Vop{\ki{}}{\Gay}\right)
+\left(\Kop{\ki{}}{\Gay}\Vop{\ke}{\Gay}+\Vop{\ki{}}{\Gay}\Kop{\ke}{*,\Gay}\right),\\
\mathcal{A}_{0,2} &= \frac{1}{2}\left(\Vop{\ke}{\Gak}-\Vop{\ki{}}{\Gak}\right)
+\left(\Kop{\ki{}}{\Gak}\Vop{\ke}{\Gak}+\Vop{\ki{}}{\Gak}\Kop{\ke}{*,\Gak}\right).
\end{eqnarray*}
Now
\begin{equation*}
\mcAe =
\begin{pmatrix}
\frac{\eeta}{2}\Vop{\ki{}}{\Gay} - \eeta\Vop{\ki{}}{\Gay}\Kop{\ke}{*,\Gay} & 0\\
0 & \frac{\eeta}{2}\Vop{\ki{}}{\Gak} - \eeta\Vop{\ki{}}{\Gak}\Kop{\ke}{*,\Gak}
\end{pmatrix}
+ \mcAo
\end{equation*}
which is of the form assumed in Lemma \ref{lemma412},
\begin{equation}
\mcAe = \eeta \mathcal{A}_1 + \mcAo,
\end{equation}
where 
\[
 \mathcal{A}_1 = \begin{pmatrix}
\frac{1}{2}\Vop{\ki{}}{\Gay} - \Vop{\ki{}}{\Gay}\Kop{\ke}{*,\Gay} & 0\\
0 & \frac{1}{2}\Vop{\ki{}}{\Gak} - \Vop{\ki{}}{\Gak}\Kop{\ke}{*,\Gak}
\end{pmatrix}.
\]
For the off-diagonal, infinitely smoothing part, we have an analogous decomposition
\begin{equation}
\mathcal{M}_\eeta = \eeta \mathcal{M}_1 + \mathcal{M}_0.
\end{equation}
Consider the difference
\begin{equation*}
\Phi(\eeta) := \pyve{\phi_1(\eeta)}{\phi_2(\eeta)} - \pyve{\phi_1}{\phi_2} .
\end{equation*}
Then
$$
\Phi(\eeta) = \left[(\AMe)^{-1} - (\AMo)^{-1}\right]\tilde{f}
$$
and thus
$$
(\AMo)\Phi(\eeta) = \left[(\AMo)(\AMe)^{-1}-id\right]\tilde{f}.
$$
We now can apply Lemma \ref{lemma412} with $m=-1$ and
\begin{eqnarray*}
P_0(x,D) &= \AMy \quad (\text{ of order }-1)\\
P_1(x,D) &= \AMo \quad(\text{ of order }-2)
\end{eqnarray*}
and hence, as  $\eeta \to 0$ outside a conical neighborhood $\Sigma$ of $i\RR$,
$$
\normi{(\AMo)\Phi(\eeta)}_{\Sob{s}{\Gay}\times\Sob{s}{\Gak}}
\leq C, \quad \eeta \to 0, \quad \eeta \notin \Sigma, 
$$
with \(C\) independent 0f \(\eeta\). Since $\AMo$ elliptic of order $-2$, we also have
$$
\normi{\Phi(\eeta)}_{\Sob{s-2}{\Gay}\times\Sob{s-2}{\Gak}}
\leq \tilde{C}, \quad \eeta \to 0, \quad \eeta \notin \Sigma
$$
and hence by compact embedding (given $\lambda>0$) there exists a  subsequence $\eeta_\nu \to 0$, $\eeta_\nu \notin \Sigma$, s.t.
$$
\normi{\Phi(\eeta_\nu)}_{\Sob{s-2-\lambda}{\Gay}\times\Sob{s-2-\lambda}{\Gak}}
\xrightarrow[\nu\to\infty]{} 0.
$$
This finishes the proof of ~\ref{prop411}. \($\Box$\)

\bigskip

\section{Presence of w--ALR}

In this section we consider dimension $d\geq 2$ and frequencies $\omega\geq 0$.
We denote by $\aepsilon_\eeta(x)$  the piecewise constant function in $\R^d$ that
is $\aepsilon_\eeta(x)=\ee$ for $x\in (\R^d\setminus \Omega)\cup D$
and $\aepsilon_\eeta(x)=
 \ee (-1+ \eeta)$ for $x\in \Omega\setminus D$, $\eeta\in i\R_+\cup\{0\}$.   
 Also,  $b=-1$ in equation (\ref{div type eq.}).
 
 \begin{figure}[htbp]
\begin{center}

\psfrag{1}{$z$}
\psfrag{2}{$S_+$}
\psfrag{3}{$S_-$}
\psfrag{4}{$\Omega$}
\psfrag{5}{$D$}
\includegraphics[width=7.5cm]{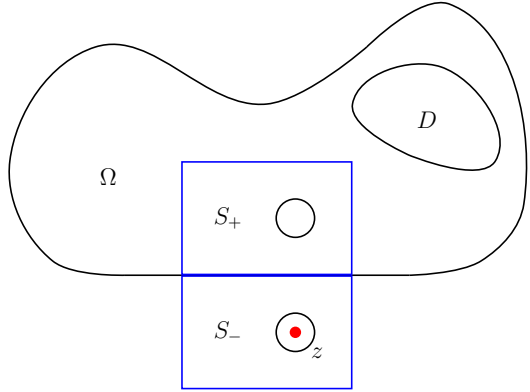}
\end{center}
\caption{Setting of the Theorem 5.1: Domain $\Omega\subset \R^d$ that contains domain $D$.
The material parameters approach in the set $\Omega\setminus D$ negative value
and are positive outside this set.
}\label{Fig-2}
 \end{figure}

\begin{theorem}\label{thm: blow up thm}

Assume $D\subset \Omega\subset \R^d$, $d\geq 2$ that the interfaces  \(\Gamma _1=\partial D \) and  \(\Gamma _2=\partial \Omega\) are smooth and 
$\Gamma_2$ contains a flat subset  $S_0=\{y_1\}\times B$, 
where $y_1\in \R$ and $B=\{x'\in \R^{d-1};\ |x'|<R_0\}$.
Also, assume that $S_+\subset \Omega\setminus \overline D$
and $S_-\subset \R^d\setminus \overline \Omega$, where
$S_+=(y_1,y_1+a)\times B$ and $S_-=(y_1-a,y_1)\times B$, $a>0$. 

Moreover, let
 \(f=\delta_z\) with $z\in S_-$.
Also, let $\tau _\eeta  = -1  + \eeta$, $\eeta\in i\R_+$, and assume that $\ke =\ki\in \R_+\cup\{0\}$,
i.e., $b=-1$ in equation (\ref{div type eq.}), and 
(\ref{Matti 1}), (\ref{Matti 2}), and (\ref{Matti 3}) are valid.
Let \(0<|\eeta| \leq \eeta _0\) for some positive fixed \(\eeta _0\). Assume the problem \HeSo with $\tau  = \tau_\eeta$ is uniquely solvable  and  that $v_i^\eeta$, $i=1,2,3$ are its solutions and \(v_i\), \(i =1,2,3\)  the solutions given by Theorem \ref{tokathm}.
Let $r_1>0$ be such that $B(z,r_1)\subset S_+$.
Then as $\eeta \to 0$,
$$
\limsup_{\eeta\to 0}\|v_2^\eeta\| _{H^{1}(S_+)}+\|v_3^\eeta\| _{H^{1}(S_-\setminus  B(z,r_1))}=\infty .
$$
\end{theorem}

\proof 
Let $S=S_+\cup S_0\cup S_-$ and let $r_2>r_1$ be such that $B(z,r_2)\subset S_-$.
Let $z^-=z=(z_1,z')\in S_-$ and $z^+=(2y_1-z_1,z')\in S_+$. 
By (\ref{reunab}), we see that there are functions
 $w^\eeta\in H^1(S)$ such that $ w^\eeta|_{S_+}=v_3^\eeta|_{S_+}$ and
  $ w^\eeta|_{S_-}=v_2^\eeta|_{S_-}$.
 To show the claim,  assume the opposite: We assume that there is a sequence
$\eeta_j\to 0$   such that $
\|w^{\eeta_j}\| _{H^{1}(S)}$ is bounded
by some constant $C_0$. 
Using \cite[Thm. 8.8]{GT} and equation (\ref{Helmholtz}), we see that there is $C_1>0$
such that  the norm of $w^{\eeta_j}|_{B(z^+,r_2)\setminus \overline B(z^+,r_1)}$  
in $H^2(B(z^+,r_2)\setminus  \overline B(z^+,r_1))$
are bounded by $C_1$   and  
the norm of $w^{\eeta_j}|_{B(z^-,r_2)\setminus  \overline B(z^-,r_1)}$  
in $H^2(B(z^-,r_2)\setminus  \overline B(z^-,r_1))$
are bounded by $C_1$.

Then by replacing $\eeta_j$ by
a suitable subsequence, that we continue to denote by $\eeta_j$,
we can assume that $w_j$ converge weakly in $H^{1}(S)$ to some
function $W$,
the restrictions  $w^{\eeta_j}|_{B(z^+,r_2)\setminus  \overline B(z^+,r_1)}$  
converge weakly to the restriction of $W|_{B(z^+,r_2)\setminus  \overline B(z^+,r_1)}$
in $H^2(B(z^+,r_2)\setminus  \overline B(z^+,r_1))$, and
the restrictions  $w^{\eeta_j}|_{B(z^-,r_2)\setminus  \overline B(z^-,r_1)}$  
converge weakly to the restriction of $W|_{B(z^-,r_2)\setminus  \overline B(z^-,r_1)}$
in $H^2(B(z^-,r_2)\setminus  \overline B(z^-,r_1))$.  
Then for all $\phi\in C^\infty_0(S\setminus \overline B(z^-,r_1))$ 
we have
\ba
& &\int_{\R^d}( \aepsilon_{0}(x) \nabla W\,\cdotp \nabla \phi-\omega ^2\aepsilon_0(x)\mu_0\ W\phi)dx\\
& &=\lim_{j\to \infty}
\int_{\R^d}( \aepsilon_{\eeta_j}(x) \nabla w_{\eeta_j}\,\cdotp \nabla \phi-
\omega ^2\aepsilon_{\eeta_j}(x)\mu_0\  w_{\eeta_j}\phi)dx=0.
\ea
Hence in the domain $S\setminus \overline B(z^-,r_1)$ we have in the sense of distributions
\beq\label{eq: elliptic eq}
\nabla\,\cdotp( \aepsilon_{0}(x) \nabla W) +\omega ^2\aepsilon_0(x)\mu_0\ W=0
\eeq
in weak sense. In particular, this yields that $W$ satisfies
an elliptic equation in $S_-$ and $S_+$  with  trace $W|_{S_0}\in H^{1/2}(S_0)$ and
$S_0$ and thus $W$ has well defined one-sided normal 
derivatives
on $S_0\subset \Ga_2$ that take values in $ H^{-1/2}(S_0)$. Applying integration
by parts in domains $S_-\setminus \overline B(z^-,r_1))$  and $S_+$
we obtain for 
$\phi\in C^\infty_0(S\setminus \overline B(z^-,r_1))$ 
\ba
0&=&\int_{\R^d}( \aepsilon_{0}(x) \nabla W\,\cdotp \nabla \phi-\omega ^2\aepsilon_0(x)\mu_0\ W\phi)dx\\
 &=&\int_{S_+}( \aepsilon_{0}(x) \nabla W\,\cdotp \nabla \phi-\omega ^2\aepsilon_0(x)\mu_0\ W\phi)dx\\
& &+\int_{S_-}( \aepsilon_{0}(x) \nabla W\,\cdotp \nabla \phi-\omega ^2\aepsilon_0(x)\mu_0\ W\phi)dx\\
\\
&=&-\int_{\p S_+}(\nu\,\cdotp \aepsilon_{0}\nabla W|_{\p S_+})  \phi \,dS(x)\\
& &+\int_{\p S_-}(\nu\,\cdotp \aepsilon_{0} \nabla W|_{\p S_-}) \phi \,dS(x)
\\
&=&-\int_{S_0}(\nu\,\cdotp  \nabla W|_{S_0-}+
\nu\,\cdotp  \nabla W|_{S_0+} ) \phi \,dS,
\ea
where $\nu=(0,0,\dots,0,1)$ is normal vector of $S_0$ pointing from \(S_-\) to \(S_+\). Thus we see that 
$\donu W\vert_{S_0-} = - \donu W\vert_{S_0+}$ and summarizing, we have
\begin{equation}\label{reunab for W}
W\vert_{S_0-} = W\vert_{S_0+}, \quad \donu W\vert_{S_0-} = - \donu W\vert_{S_0+}.
\end{equation} 
Using this, equation (\ref{eq: elliptic eq}),   and the fact that $S\setminus  (\overline B(z^-,r_1)\cup \overline B(z^-,r_1))$
is connected implies that  for $x=(x_1,x')\in F:=
S\setminus  (\overline B(z^-,r_1)\cup \overline B(z^+,r_1))$ we have the symmetry
\beq\label{symmetry}
W(x_1,x')=W(2y_1-x_1,x'),\quad x=(x_1,x')\in F.
\eeq
Then we see using the Gauss theorem that
\begin{equation}
\int_{\p B(z^-,r_1)} \p_\nu  w_{\eeta_j}(x)dS(x) = 1 -  \omega ^2\mu _0 \int_{B(z^-,r_1)} w_{\eta _j}\, dx.
\end{equation}
On the other hand,
\ba
& &  \lim_{j\to \infty}\int_{\p B(z^-,r_1)} \p_\nu  w_{\eeta_j}(x)dS(x)\\
&=&\int_{\p B(z^-,r_1)} \p_\nu  W(x)dS(x)\\
&=&\int_{\p B(z^+,r_1)} \p_\nu  W(x)dS(x)\\
&=&\lim_{j\to \infty}\int_{\p B(z^+,r_1)} \p_\nu  w_{\eeta_j}(x)dS(x)\\
&=&0.
\ea
Hence, for all \(r_1 >0\),
\[
1 =  \omega ^2\mu _0 \lim_{j\to \infty} \int_{B(z^-,r_1)} w_{\eta _j}\, dx.
\]
Now the sequence \((w_{\eta_j})\) was bounded in \(H^1 (S)\), so that
\[
\bigg|\int_{B(z^-,r_1)} w_{\eta _j}\, dx\bigg| \leq Cr_1^{d/2},
\] 
which yields a contradiction as \(r_1 \to 0\). \qed

\medskip

 The results of the previous sections, together with the earlier  results of Milton et al. 
 ( \cite{Milton05,Milton06})
and Ammari et al. (\cite{ACKLM}) show that  for strictly convex bodies ALR
may  appear only for bodies so small that the quasi-static approximation is realistic.
This gives limits for size of the objects for which invisibility cloaking methods based on ALR
may be used. However, 
 the results of this section show that
  the weak  ALR may appear if the body $\Omega\setminus \overline D$ has double negative material parameters
  and its external boundary contains flat parts.

\section*{Aknowledgments}
This work was supported by the Academy of Finland (Centre of Excellence in Inverse Problems Research 2012--2017 and LASTU research program through COMMA project).

\section*{Appendix: Anomalous localized resonance in electromagnetics and numerical examples}

Let us consider a time-harmonic TE$_z$-polarized electromagnetic wave propagating in the $xy$-plane in source-free space. Let \(\vr = (x,y,z)\) be the position vector in euclidean coordinates. The electric and magnetic fields can then be written as
\[
\vek{E}(\vr) = \uvx E_x(x,y) + \uvy E_y(x,y), \, \, 
\vek{H}(\vr) = \uvz H_z(x,y),
\]
where $u_x,u_y$, and $u_z$ are the unit coordinate vectors.
Assuming the background medium isotropic but inhomogeneous, the Maxwell equations become
\beq
\nabla \times \vek{E}(\vr) &=& i\omega\mu_0\mur(\vr)\vek{H}(\vr) \label{Fara} \\
\nabla \times \vek{H}(\vr) &= &-i\omega\epz\epr(\vr)\vek{E}(\vr) \label{Ampe}
\eeq
Faraday's law (\ref{Fara}) gives
\begin{equation}
\osder{}{x}E_y - \osder{}{y}E_x = i\omega\mu_0\mu_r(\vr)H_z \label{Fara2}
\end{equation}
and from Amp\`ere's law (\ref{Ampe}), 
\begin{equation}
\uvx\osder{}{y}H_z - \uvy\osder{}{x}H_z = -i\omega\epz\epr(\vr)(\uvx E_x + \uvy E_y),
\end{equation}
we can solve
\begin{equation}
E_x = \frac{i}{\omega\epz\epr(\vr)}\osder{}{y}H_z, \quad
E_y = -\frac{i}{\omega\epz\epr(\vr)}\osder{}{x}H_z.
\end{equation}
Substituting these into (\ref{Fara2}), we obtain a scalar equation for $H_z$ as
\begin{equation}
\left(\nabla \cdot \frac{1}{\epr(\vr)}\nabla + \mu_r(\vr)k_0^2\right)H_z = 0, \label{Helmholtz B}
\end{equation}
where $k_0^2 =\omega^2\epz\mu_0$.
\medskip

Let us further consider a case when this wave interacts with an infinitely long layered circular cylindrical structure where an inner core with radius $r_c$ is covered by an annular shell with radius $r_s$. Assume the structure is non-magnetic, $\mur(\vr)=1$, and the relative permittivity is given as
\begin{equation}
\epr(\vr) = \epr(r) =
\begin{cases}
1,& r > r_s\\
{\newtext-1 +i\sigma},& r_c < r < r_s\\
1,& r< r_c
\end{cases} \label{epr}
\end{equation}

It was first shown in \cite{Nicorovici94} using a quasi-static analysis at the limit $\omega \to 0$ that with vanishing material losses $\sigma \to 0$ this structure supports so called anomalous localized resonance (ALR) that can be excited by an external line dipole source with dipole moment $\vp = \uvx p_x + \uvy p_y$ located at the distance $r_0$ from the cylinder axis. Later it was shown that the structure can actually be interpreted as a cylindrical superlens \cite{Milton05}. As the permittivity of the core is chosen $\epr = 1$ being equal to the permittivity of the exterior, Theorem 3.2 of \cite{Milton05} states that the ALR is excited when the dipole is brought within the distance $r_s < r_0 < r_*$, where $r_* = r_s^2/r_c$. This interval can further be divided into two parts. If the distance of the dipole $r_\# < r_0 < r_*$, where $r_\# = \sqrt{r_s^3/r_c}$, there are two separate resonant regions around both interfaces of the negative-permittivity annulus. The resonant 
 region around the interface between the core and the shell is located between $r_0(r_c/r_s)^2$ and $r_s^2/r_0$ and the region around the interface between the shell and the exterior between $r_0r_c/r_s$ and $r_s^3/(r_cr_0)$. This indicates that the resonance phenomenon is really localized. Outside these limits the fields are well-behaved.
\medskip

Instead, if the permittivity of the inner core deviates from the one of the exterior, the ALR can be excited from a larger distance. In our case this would mean that $\epr \neq 1$, when $r < r_c$. In this case, the critical distance of the source dipole becomes $r_\text{crit} = r_s^3/r_c^2$ \cite{Nicorovici94, Milton05}. 
\medskip

\begin{figure}[!ht]
\centering \includegraphics[width=0.45\textwidth]{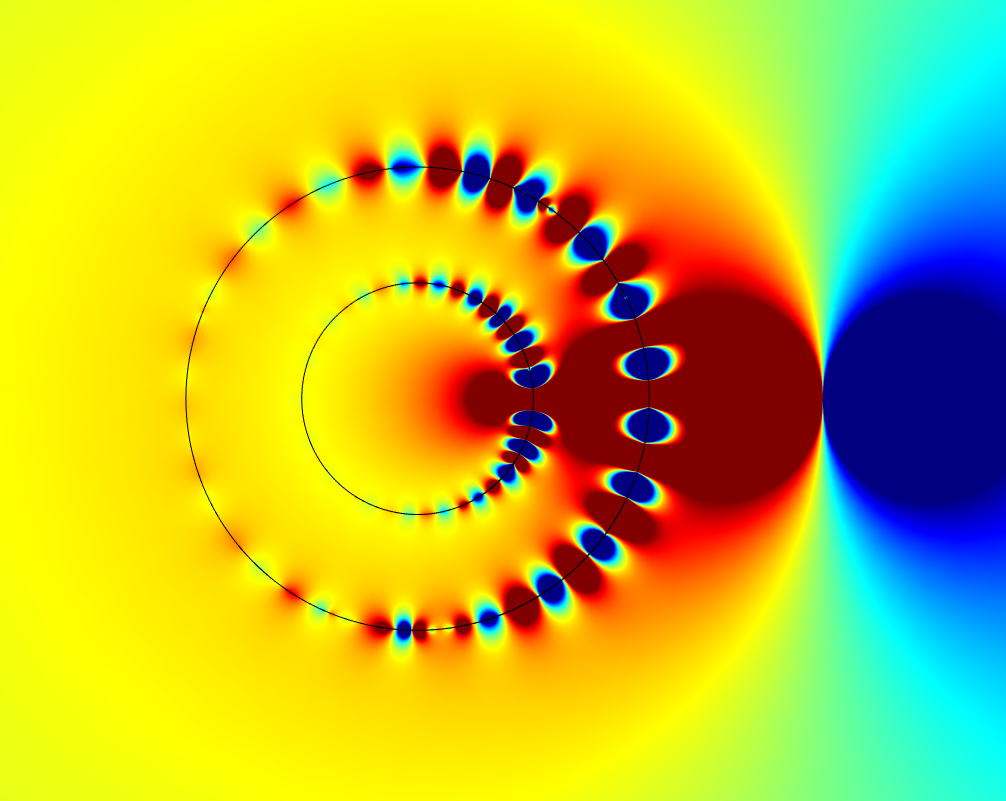} \hspace{0.5cm}
\includegraphics[width=0.46\textwidth]{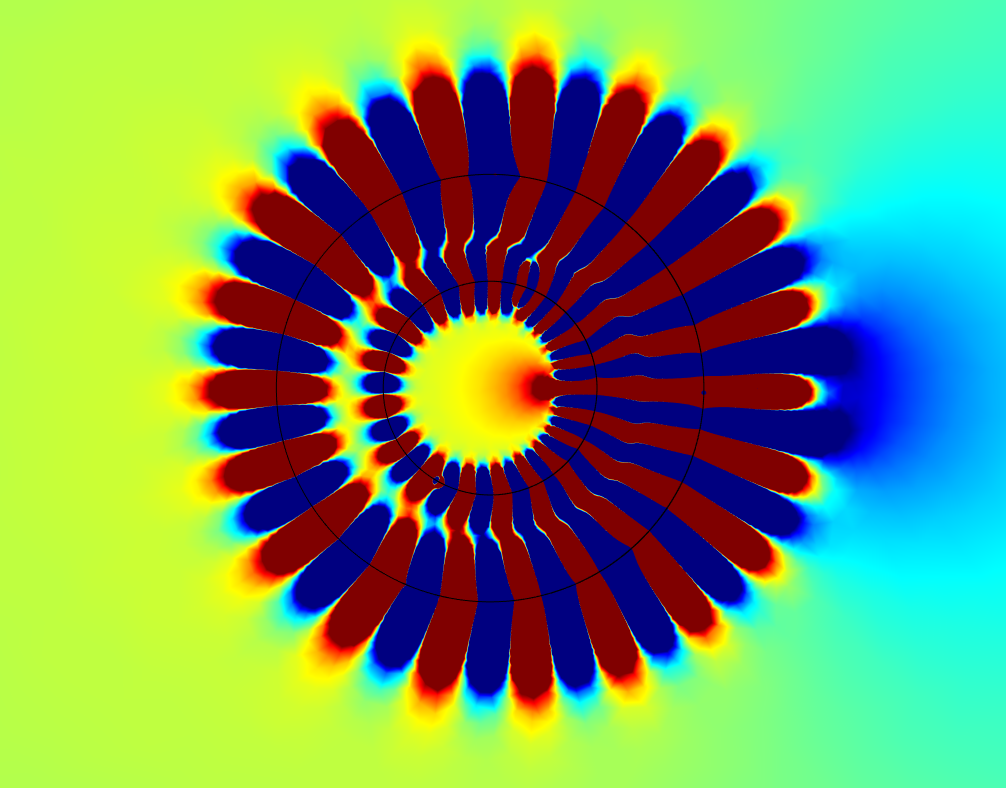}
\caption{Computational visualization of the occurance of ALR in the electrostatic potential in a quasi-static case
that has been observed and analyzed in \cite{Milton05,Milton06}. The model parameters are $r_c = 2$, $r_s = 4$, $\sigma = 1\times 10^{-7}$, $\vp = \uvx p_x$. Left panel: $r_0 = 7$, Right panel: $r_0 = 5$.} \label{ComsolMiltonExample}
\end{figure}

Let us computationally verify and visualize the quasi-static example case given in \cite{Milton05} using Comsol Multiphysics 4.4 software based on the finite element method (FEM) (see Fig.~\ref{ComsolMiltonExample}). Let us choose $r_c = 2$ and $r_s = 4$, which gives us $r_\# = \sqrt{32} \approx 5.66$ and $r_* = 8$. Let the source be a line dipole with dipole moment $\vp = \uvx p_x$ located at distance $r_0$. First the dipole is placed between $r_\#$ and $r_*$ at $r_0 = 7$. Theory suggests that there will be two separate resonant annuli, the inner one at $1.75 < r < 2.29$ and the outer one at $3.50 < r < 4.57$. The permittivity of the cylindrical structure follows Eq.~(\ref{epr}). To ensure better numerical stability, a small imaginary part $\sigma = 1\times 10^{-7}$ is added. The left panel of Fig.~\ref{ComsolMiltonExample} quite nicely agrees with the theory. However, the resonant regions are not cylindrically symmetric due to the asymmetry of the excitation. In the right 
 panel of Fig.~\ref{ComsolMiltonExample}, the source if placed between $r_s$ and $r_\#$. In this case, the outer limit of the inner resonant region, $r = 16/5 \approx 3.20$, is larger than the inner limit of the outer region $r = 2.50$. Thus the resonant annuli are overlapping and the ALR occurs within a single continuous range $1.25 < r < 6.40$. Again, the asymmetry is caused by the excitation. 
\medskip

\begin{figure}[!ht]
\centering \includegraphics[width=0.45\textwidth]{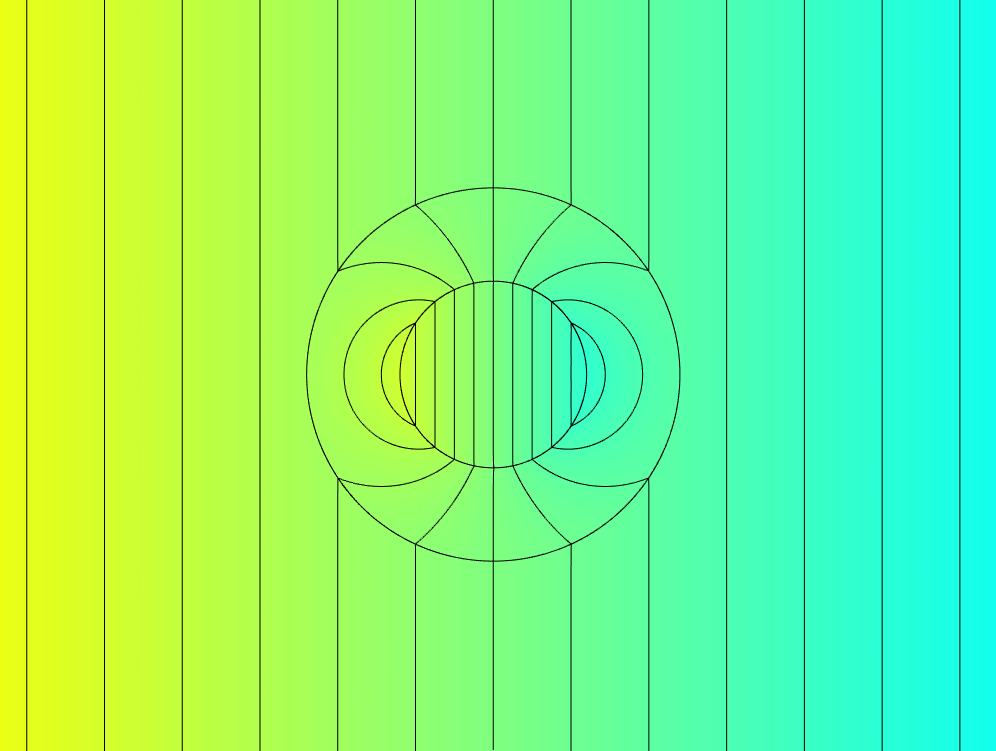} \hspace{0.5cm}
\includegraphics[width=0.46\textwidth]{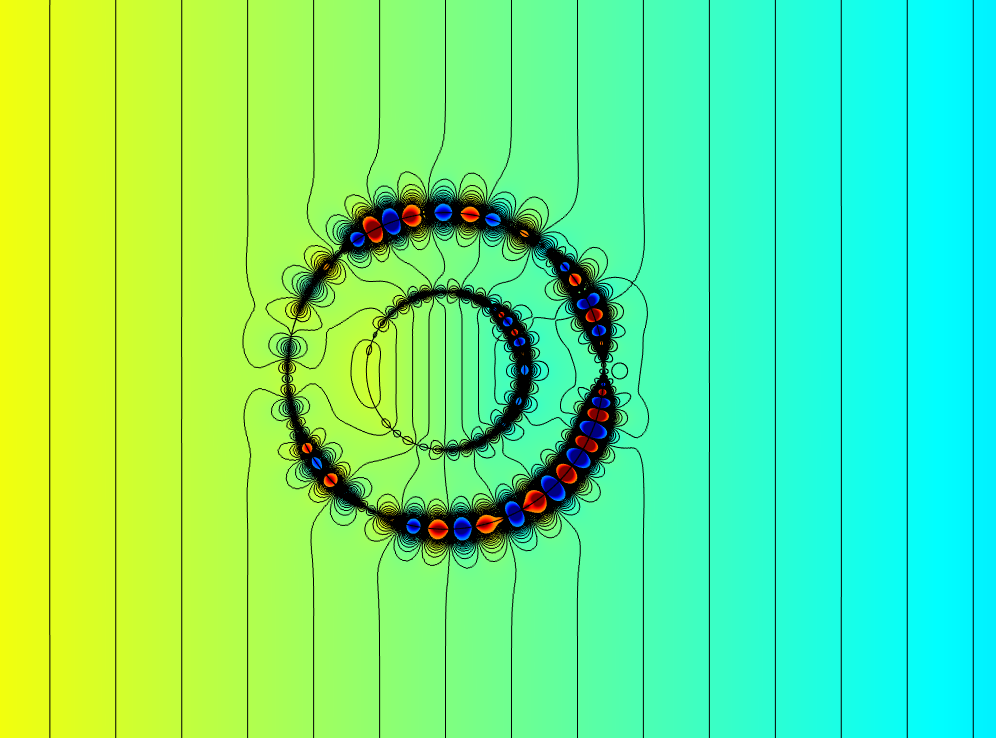}
\caption{The cloaking based on ALR with quasi-static approximation (for detailed analysis, see \cite{Milton05,Milton06} and \cite{ACKLMII}). Potential distribution in the vicinity of a layered cylindrical structure with $r_c = 1$ and $r_s = 2$. The permittivity of the annulus $r_c < r < r_s$ is $\epr = -1$. Left: The structure remains invisible in a uniform external field. Right: A small polarized cylinder at $r_0 = 2.5$ excites the ALR in the structure and becomes cloaked.} \label{ComsolCloak}
\end{figure}

Perhaps an even more fascinating detail is that the ALR can also be used for cloaking purposes \cite{Milton06}. When a polarizable line dipole is brought within the distance $r_\# = \sqrt{r_s^3/r_c}$ from the axis of the cylinder, in a uniform external electric field it becomes cloaked from an outside observer. Figure \ref{ComsolCloak} visualizes the cloaking phenomenon computationally in the quasi-static case. The left panel shows the potential distribution of the cylindrical structure with $r_c = 1$, $r_s = 2$ and permittivity $\epr(\vr)$ of Eq.~(\ref{epr}) with $\sigma = 0$ in a uniform static field. In this case, the structure causes no perturbation to the external field. In the right panel a small circular cylinder with radius $\rho_0 = 0.1$ and permittivity $\epr = 100$ is placed at the distance $r_0 = 2.5$ from the axis of the layered cylinder. In the external field, the small cylinder becomes polarized and its dipolar field excites the ALR in the layered cylindrical structure, which acts back on the small cylinder making the whole system invisible. The external field outside the distance $r_\# \approx 2.82$ remains unperturbed.  

\begin{figure}[!ht]
\centering \includegraphics[width=0.45\textwidth]{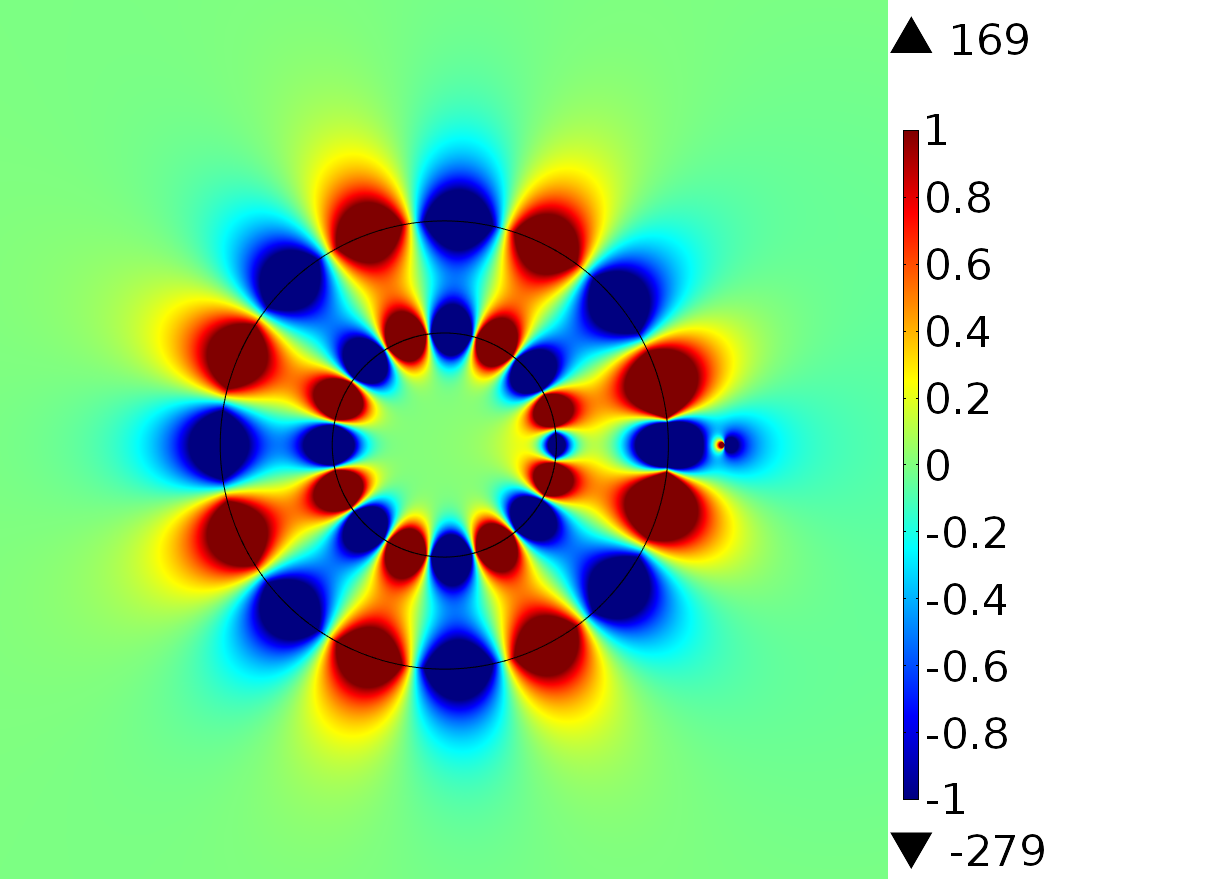} \hspace{0.5cm}
\includegraphics[width=0.45\textwidth]{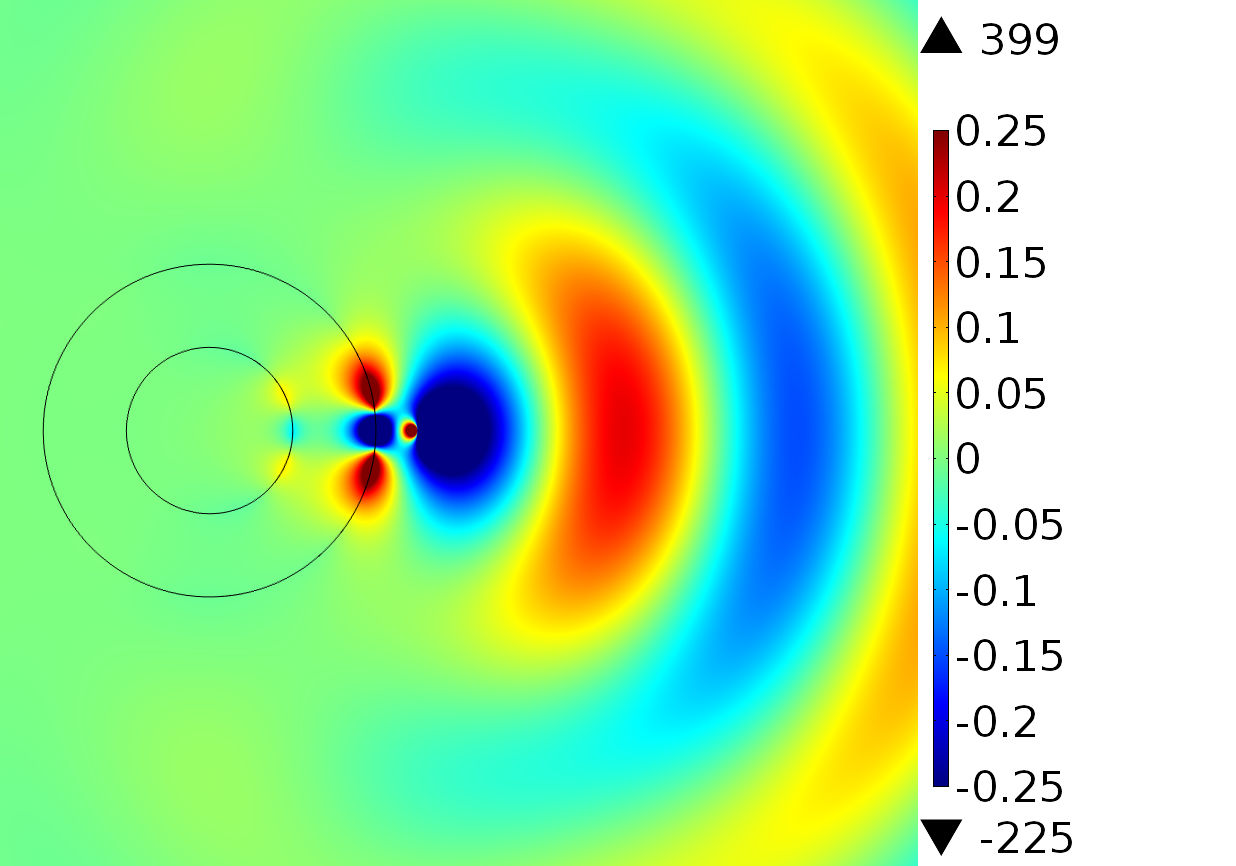}
\caption{Visualization of the axial magnetic field $H_z$ in the electrodynamic case, i.e.,
without quasi-static approximation. The model parameters are $r_c = 2$, $r_s = 4$, $\sigma = 1\times 10^{-5}$, $\vp = 1\uvy$ and $r_0 = 5$. Left panel: The frequency is $f = 12.5\times 10^6$ $(2r_s = \lambda/3)$, Right panel: The frequency is $f = 37.5\times 10^6$ $(2r_s=\lambda)$. Note, that our results do not cover the general 2D case i.e. without the quasi-static approximation. This is still an open problem.} \label{ComsolDynamic}
\end{figure} 

Next we consider  the  electrodynamic case outside the quasistatic regime where the size of the cylindrical structure is no longer significantly small compared with the wavelength. Let us computationally study the geometry setup given in the right panel of Fig.~\ref{ComsolMiltonExample} using a radiating line dipole instead of the static one
without using the quasi-static approximation. It turns out that for better numerical convergence it is more practical to choose the dipole $\uvy$-polarized and increase the material losses to $\sigma = 1\times10^{-5}$. Figure \ref{ComsolDynamic} shows the axial magnetic field component $H_z$. In the left panel, the frequency ($f=12.5\times 10^6$) is chosen such that the outer radius of the cylinder, $2r_s$, is one third of the wavelength of the fields radiated by the dipole. 
A resonance that resembles ALR is still seen occurring in the structure even though its size starts to be comparable with the wavelength. In the right panel ($f=37.5\times 10^6$) the radius of the cylinder is exactly one wavelength, $2r_s = \lambda$. Here we observe a qualitatively different behavior. Even though a very strong field enhancement is seen on the boundary of the outer annulus in the near vicinity of the dipole, the scattering from the structure is dominant and the localized resonance phenomenon that would cover the the whole structure is absent. Hence, there obviously is an upper limit for the electrical size of the cylindrical structure where the ALR type of resonance is no longer supported. With these particular geometry and material parameters, this happens approximately at $2r_s \approx 3\lambda/5.$  

Considering an actual realization is this kind of structure, a possible choice for the material of the negative-permittivity annulus could be silver at ultraviolet A range. The permittivity of silver is often described using Drude dispersion model
\begin{equation}
\varepsilon_\text{Ag}(\lambda) = \epz\left(\epsinf - \frac{(\lambda/\lambda_p)^2}{\newtext 1+i\lambda/\lambda_d}\right) \label{Drude}
\end{equation} 
where $\lambda$ denotes the free-space wavelength. Based on the measured values presented in \cite{Johnson72}, Ref. \cite{Wallen09b} applies the model (\ref{Drude}) with fitted parameters $\epsinf = 5.5$, $\lambda_p = {\rm 130\,nm}$ and $\lambda_d = {\rm 30\,nm}$ for wavelength range ${\rm 320\,nm} < \lambda < {\rm 700\,nm}$ (${\rm 430\,THz} \lesssim f \lesssim {\rm 940\,THz}$). At UVA wavelength $\lambda = {\rm 331\,nm}$ ($f \approx {\rm 906\,THz}$) this model would give silver the relative permittivity $\epr = -1 +i0.07.$
\medskip

Also, it is mentioned in \cite{Nicorovici94} that silicon carbide (SiC) would have a  relative permittivity $\epr = -1 +i 0.1$ at much longer infrared wavelength $\lambda = {\rm 10.550}\,\mu {\rm m}$ ($f \approx {\rm 28.4\,THz}$).
\medskip

\begin{figure}[!ht]
\centering \includegraphics[width=0.6\textwidth]{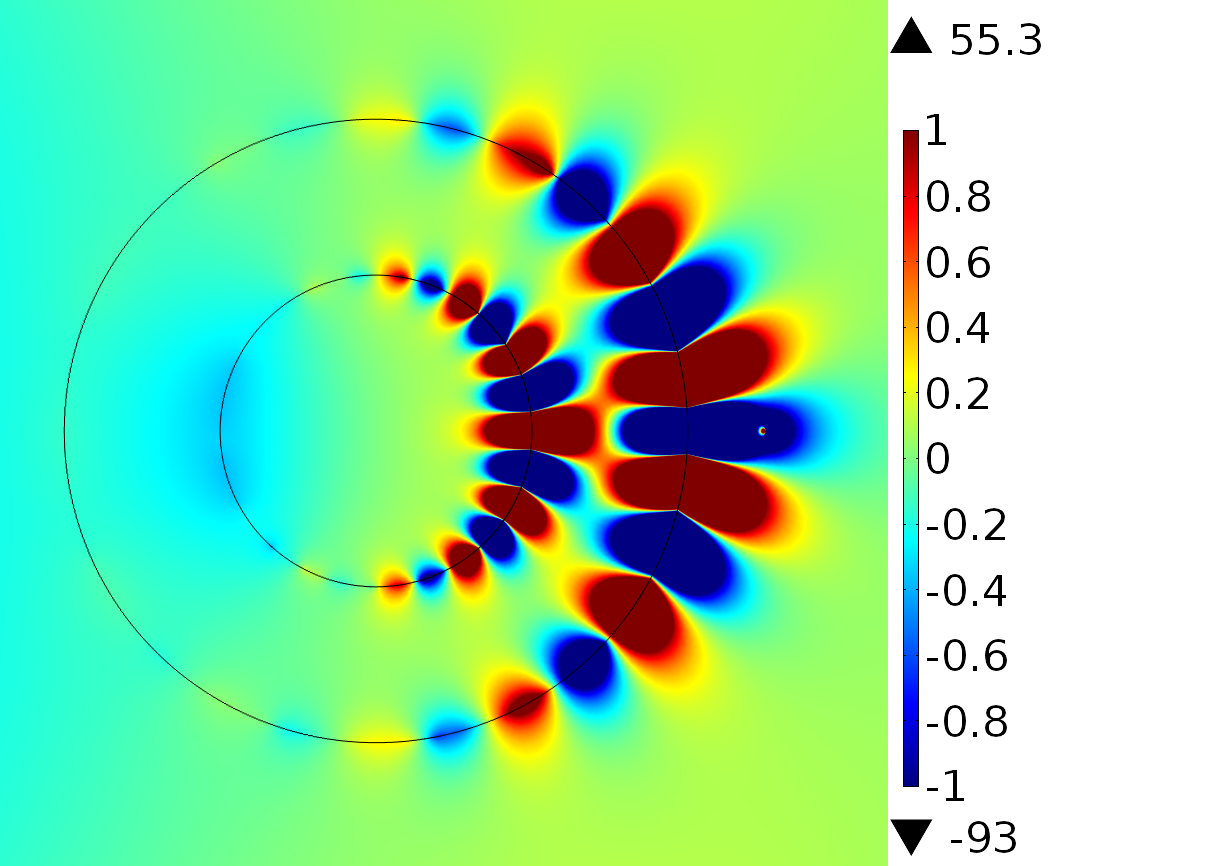} 
\caption{Visualization of the axial magnetic field $H_z$ for a double negative 2D layer in the dynamic case. Parameters: $r_c=2$, $r_s=4$, $r_0=5$, $\vp = 1\uvy$ $\epr = -1 +i1\times 10^{-5}$, $\mur = -1$, $f = 12.5\times 10^6$.} \label{DNG}
\end{figure} 

\begin{figure}[!ht]
\centering \includegraphics[width=0.6\textwidth]{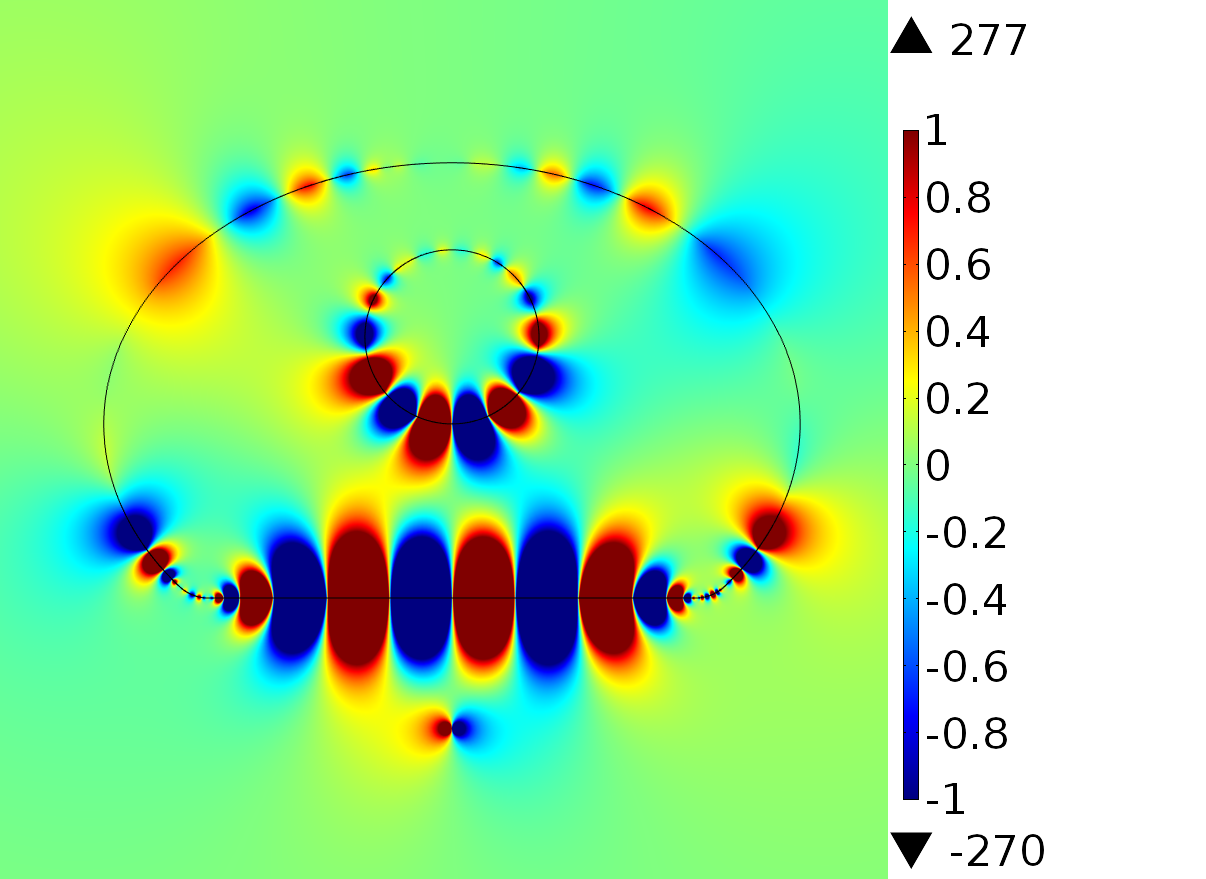} 
\caption{Visualization of the axial magnetic field $H_z$ for a 2D double negative domain with a flat part on the boundary. An ellipse semiaxes $a_s = 4$ and $b_s = 3$, material parameters $\epr = -1 + i1\times 10^{-5}$, $\mu = -1$ and a circular hole with radius $r_c = 1$ is excited by a dipole $\vp_y = 1\uvy$ at frequency $f=1.25\times 10^6$. The lower part of the ellipse is cut flat, and the resulting corners have been smoothed out. The dipole is located $r_0 = 3.5$ below the center point of the ellipse.} \label{DNGflat}
\end{figure}

Above it is assumed that only the permittivity of the annulus is negative. In the lossless case with $\epr = -1$ and $\mur = 1$, the wave number $k = k_0\sqrt{\epr}\sqrt{\mur}$ becomes purely imaginary and no wave propagation inside the annulus is possible. Also, $k^2 = -k_0^2$ and the Helmholtz equation (\ref{Helmholtz B}) inside the annulus $r_s < r < r_c$ reduces to the form
\begin{equation}
(\Delta - k_0^2)H_z = 0.
\end{equation}

Instead, if also $\mur = -1$, the material of the annulus becomes double negative and supports propagating backward waves with negative wave number $k = k_0\sqrt{\epr}\sqrt{\mur} = -k_0$. The square of $k$ is again positive resulting into the ordinary Helmholtz equation
\begin{equation}
(\Delta +  k_0^2)H_z = 0.
\end{equation}

Figure \ref{DNG} visualizes the case for a double negative annulus. The parameters are otherwise the same as in the left panel of Fig.~\ref{ComsolDynamic}, except for $\mu = -1$. We note that in this double negative case, the resonance does not occur symmetrically around the structure, but is more focused in the vicinity of the exciting dipole. Furthermore, the maximum amplitude of the field has decreased. In Fig.~\ref{DNGflat}, the structure is reshaped to resemble the one depicted in Fig.~\ref{Fig-2}. The shape of the outer annulus is elliptic and the center of the inner circular hole is located $0.5$ above the center of the ellipse. Furthermore, the bottom of the ellipse is cut flat, and the resulting corners have been rounded to ensure that the interface remains smooth. The dipole with $\vp = 1\uvy$ is now located $3.5$ below the center of the ellipse. The material parameters and the frequency are the same as in Fig.~\ref{DNG}. We note that the strongest resonance is focused on the flat part of the interface.

\end{document}